\documentclass[aps,twocolumn,prc,floatfix,showpacs]{revtex4-1}
\usepackage{epsf,color,colordvi}
\usepackage{graphicx}
\usepackage{amsmath,amssymb}
\usepackage{times}
\usepackage{subfigure}
\usepackage{todonotes}
\usepackage{pstricks}
\usepackage{placeins}
\usepackage{ulem} 
\renewcommand{\theequation}{\arabic{section}.\arabic{equation}}

\begin{document}
\title{Numerical solutions of the Schr\"odinger equation with source terms or time-dependent potentials}

\author{W. van Dijk}
\affiliation{Department of Physics, Redeemer University College, Ancaster, Ontario L9K 1J4, Canada}
\affiliation{Department of Physics and Astronomy,
McMaster University, Hamilton, Ontario L8S 4M1, Canada}
\email{vandijk@physics.mcmaster.ca}

\author{F. M. Toyama}
\affiliation{Department of Computer Science, Kyoto Sangyo University, Kyoto 603-8555, Japan}

\begin{abstract}
We develop an approach to solving numerically the time-dependent Schr\"odinger equation when it includes source terms and time-dependent potentials.  The approach is based on the generalized Crank-Nicolson method supplemented with an Euler-MacLaurin expansion for the time-integrated nonhomogeneous term.  By comparing the numerical results with exact solutions of analytically solvable models, we find that the method leads to precision comparable to that of the generalized Crank-Nicolson method applied to homogeneous equations.  Furthermore, the systematic increase in precision generally permits making estimates of the error.
\end{abstract}

\date{\today} 

\maketitle
\section{Introduction}
\label{sec:1}
Recent interest in accurate numerical solutions of the time-dependent Schr\"odinger equation (TDSE) using a generalized Crank-Nicolson (CN) approach~\cite{vandijk07,wang09}, suggests further study for cases where the Schr\"odinger equation has a nonhomogeneous term and  where the Hamiltonian is time dependent.  Over the years the method of choice for solving the homogeneous Schr\"odinger equation with time-independent interactions has been the Chebyshev expansion of the propagator introduced in 1984 by Talezer and Kosloff~\cite{talezer84}.  However, comparison of this method with the generalized CN approach, in which the time-evolution operator is expressed as a Pad\'e approximant~\cite{vandijk11}, demonstrates that the two approaches are similar in efficiency and accuracy.  Under different circumstances either method may outperform the other~\cite{formanek10}.  One advantage of the Pad\'e approach is that it is explicitly unitary, whereas the Chebyshev-expansion approach is not.  Since the latter can give very precise wave functions, however, this does not seem to be an issue in practice.

The Chebyshev-propagator method has recently been applied to nonhomogeneous Schr\"odinger equations~\cite{ndong09} and to time-dependent Hamiltonians~\cite{ndong10}.  
Given that Pad\'e-approximant expression of the propagator yields comparable results for homogeneous systems, this paper explores the extension of the Pad\'e-approximant method to solving nonhomogeneous equations. A natural sequitur is an approach to solve equations in which the interaction is time-dependent.  For such a case the time-dependent interaction term can be considered to be the nonhomogeous term and  the solution can be obtained by self-consistent iterations.  We also discuss this approach.  A decided advantage of the method discussed in this paper is that the basic calculations are unitary whereas the wavefunctions do not in general have time-independent normalization.  The calculations with unitary operators places a strong constraint on the problem resulting in stable solutions. 

Solutions of the TDSE form the basis of the  study of a multitude of nonrelativistic quantum systems.  For stationary  states, such as bound states, one can reduce the problem to the determination  of solutions of the time-independent Schr\"odinger equation.  For detailed investigations of quasi-stable systems or more general time-dependent systems one needs to solve the TDSE.  There are only a few analytically solvable models (see, e.g., Refs.~\cite{vandijk99,vandijk02} and references contained in them), but most realistic systems require numerical solutions.  In an earlier paper~\cite{vandijk07} (hereafter referred to as I) we presented an accurate and efficient method for obtaining solutions of the homogeneous Schr\"odinger equation in one dimension and for uncoupled partial waves in three dimensions.  

In some problems, however, it is necessary to solve the nonhomogeneous
 Schr\"odinger equation.  Among others, two important classes of problems involve such equations.  The first concerns systems in which the Hamiltonian can be split into parts, one of which leads to an exact analytic solution. Consider the Hamiltonian of a system $H = H_0 + V_1$.  The wave function $\Psi$ describing the system is the solution of
\begin{equation}\label{eq:0.1}
\left(i\hbar\dfrac{\partial }{\partial t} - H\right)\Psi = 0.
\end{equation}
If the wave function of the system with $H_0$ instead of $H$ is $\Psi_0$, we can obtain $\Psi$ through a correction $\Psi_1$, so that 
$\Psi=\Psi_0+\Psi_1$, by solving
\begin{equation}
\left(i\hbar\dfrac{\partial}{\partial t} -H_0 - V_1\right)
\Psi_1 = V_1\Psi_0,
\end{equation}
where $\left(i\hbar\dfrac{\partial}{\partial t}-H_0\right)\Psi_0 =0$.
This formulation is exact and may also be useful when $\Psi_0$ is known analytically and $V_1$ not necessarily small.

The second class deals with problems associated with reactions in which particles are created or annihilated.  The nonhomogeneity in the TDSE  plays the role of a source or sink of these particles.  The bremsstrahlung associated with $\alpha$ decay is an example of such a process~\cite{nogami01,vandijk03a}.

The interaction of particles with a strong radiation field can be formulated in terms of a TDSE in which the Hamiltonian is explicitly time-dependent~\cite{ndong10,muller99a}.  Such systems can be formulated as nonhomogenous equations where the wave function is  a factor in the source term.  The solution for the nonhomogeneous equation can be adapted to solve such equations.
 
In this paper we present a method of numerically obtaining solutions to the nonhomogeneous Schr\"odinger equation which are accurate to an arbitrary order of the spatial and temporal step size.  The method, like that for the homogeneous Schr\"odinger equation~\cite{vandijk07}, proves to be capable of high precision and 
efficiency.  

In Sec.~\ref{sec:2} we derive the numerical solution to the nonhomogeneous equation.  We do this in stages to develop the notation and eventually generalize the method to arbitrary order in time. The approach is evaluated by comparison to analytically known solutions in Sec.~\ref{sec:3}.  In Sec.~\ref{sec:4} the numerical solutions when the interaction depends on time is discussed and compared to known exact solutions. We conclude with summary comments in Sec.~\ref{sec:5}.

\vspace{\baselineskip}
 
\section{Generalized Crank-Nicolson method in the presence of a
  nonhomogeneous term}
\label{sec:2}  
\setcounter{equation}{0}
Let us consider the TDSE with a
nonhomogenous term.  Suppressing the dependence on spatial
coordinate(s) we write the equation as
\begin{equation}\label{eq:1.1}
  \left(i\hbar\frac{\partial~}{\partial t} - H\right)\psi(t) = N(t).
\end{equation}
For now we assume that the Hamiltonian $H$ is independent of time $t$.  The
homogeneous equation corresponding to Eq.~(\ref{eq:1.1}) has a
solution which can be written in terms of the time-evolution operator,
i.e.,
\begin{equation}\label{eq:1.2}
  \psi_\mathrm{h}(t+\Delta t) = e^{-iH\Delta t/\hbar}\psi_\mathrm{h}(t) 
\end{equation}
The nonhomogeneous equation has a particular solution
\begin{equation}\label{eq:1.3}
  \psi_\mathrm{nh}(t+\Delta t) = -\frac{i}{\hbar}e^{-iH(t+\Delta t)/\hbar}
  \int_t^{t+\Delta t} e^{iHt'/\hbar} N(t') \ dt'.
\end{equation}
The general solution is 
\begin{equation}\label{eq:02.6}
\psi(t) = \psi_\mathrm{h}(t) + \psi_\mathrm{nh}(t)
\end{equation}
with the boundary condition value inserted such that $\psi(t_0) =
\psi_\mathrm{h}(t_0) = \phi$ where $\phi$ is a normalized function of the
spatial coordinate(s).  Thus the solution with the \textit{appropriate boundary condition} may be obtained by increasing $t$ (starting at $t_0$) by steps equal to $\Delta t$ using
\begin{equation}\label{eq:1.4}
\begin{split}
  \psi(t+\Delta t) = & e^{-iH\Delta t/\hbar}\psi(t) \\
  & - \frac{i}{\hbar} 
  e^{-iH\Delta t/\hbar}
  \int_0^{\Delta t} e^{iH\theta/\hbar} N(t+\theta) \ d\theta.
  \end{split}
\end{equation}

\begin{widetext}
\subsection{Trapezoidal Rule}
Using the trapezoidal rule for the integral in Eq.~(\ref{eq:1.4}), we obtain 
\begin{equation}\label{eq:1.5}
  \psi(t+\Delta t) = e^{-iH\Delta t/\hbar}\psi(t) -\frac{i}{\hbar}
  e^{-iH\Delta t/h}
  \frac{\Delta t}{2} \left[e^{iH\Delta t/\hbar}N(t+\Delta t) + N(t)\right] 
  + {\cal O}((\Delta t)^3).
\end{equation}
In the spirit of Moyer~\cite{moyer04}, we write
\begin{equation}\label{eq:1.6}
  \psi(t+\Delta t) + \frac{i\Delta t}{2\hbar} N(t+\Delta t) = e^{-iH\Delta t/\hbar}
  \left(\psi(t) - \frac{i\Delta t}{2\hbar}N(t)\right)+ {\cal O}((\Delta t)^3).
\end{equation} 
Expanding the time evolution operator to the lowest-order unitary form, we obtain
\begin{equation}\label{eq:1.7}
  \psi(t+\Delta t) + \frac{i\Delta t}{2\hbar} N(t+\Delta t) = 
  \frac{1-\frac{\textstyle i}{\textstyle 2\hbar}H\Delta t}{1+
    \frac{\textstyle i}{\textstyle 2\hbar}H\Delta t}
  \left(\psi(t) - \frac{i\Delta t}{2\hbar}N(t)\right)+ {\cal O}((\Delta t)^3).
\end{equation}
The expansion of the time-evolution operator and the trapezoidal rule
both give an error term that is of  third order  in $\Delta t$.  We
rewrite this equation as
\begin{equation}\label{eq:1.8}
  \left(1+\frac{i}{2\hbar}H\Delta t\right)\left[\psi(t+\Delta t)
    +\frac{i\Delta t}{2\hbar}N(t+\Delta t)\right] = 
  \left(1-\frac{i}{2\hbar}H\Delta t\right)\left[\psi(t)
    -\frac{i\Delta t}{2\hbar}N(t)\right]+ {\cal O}((\Delta t)^3).
\end{equation}
If we include the $x$ dependence of $\psi(t)$ and $N(t)$ explicitly,
the equation is
\begin{equation}\label{eq:1.9}
  \left(1+\frac{i}{2\hbar}H\Delta t\right)\left[\psi(x,t+\Delta t)
    +\frac{i\Delta t}{2\hbar}N(x,t+\Delta t)\right] = 
  \left(1-\frac{i}{2\hbar}H\Delta t\right)\left[\psi(x,t)
    -\frac{i\Delta t}{2\hbar}N(x,t)\right]+ {\cal O}((\Delta t)^3),
\end{equation}
\end{widetext}
and is similar to Eq.~(2.5) of I.  It can therefore be solved numerically as outlined
in Sec. II of I to any order of accuracy in $\Delta x$.  We define
\begin{equation}\label{eq:1.10}
  \Psi^{(\pm)}(x,t) = \psi(x,t) \pm \frac{i\Delta t}{2\hbar}N(x,t).
\end{equation}
The solution with a time advance of step $\Delta t$ is found by
solving the equivalent of Eq.~(2.12) of I, i.e.,
\begin{equation}\label{eq:1.11}
  A\Psi^{(+)}_{n+1} = A^*\Psi^{(-)}_n,
\end{equation}
where the matrix $A$ is defined in I and the vector $\Psi_n^{(\pm)}$ has
components $\psi_{j,n}\pm (i\Delta t/2\hbar)N_{j,n}$.  (As in I we use partitions of $x$: $x_0,x_1,
\dots,x_j,\dots,x_J$ with $\Delta x=x_j-x_{j-1}$ and of $t$:
 $t=t_0,t_1,\dots, t_n,\dots$ with $\Delta t= t_n-t_{n-1}$.)  
Since $N(x,t)$ is a given known
function for all $x$ and $t$, and $\psi(x,t)$ is presumed known from
the calculation of the previous step, $\psi(x,t+\Delta t)$ can be
determined from the calculated $\Psi^{(+)}_{n+1}$.  Thus we obtain a
solution which has an error of ${\cal O}((\Delta x)^{2r})$ for any
integer $r>0$ in the $x$ dependence and of ${\cal O}((\Delta t)^3)$ in
the $t$ dependence.  (The parameter $r$ determining the order of the spatial integration is defined in I.) 

\subsection{Improved integration over time}

In order to obtain higher order approximations to the time evolution
of the solution of the nonhomogenous TDSE, we use a quadrature of
higher order than the trapezoidal rule in Eq.~(\ref{eq:1.4}).  Let us consider the
Euler-MacLaurin formula~\cite[formula 23.1.31]{abramowitz65},
\begin{widetext}
\begin{eqnarray}\label{eq:1.12}
  \int_0^{\Delta t} f(\theta) \ d\theta & = &
  \frac{\Delta t}{2}[f(\Delta t)+f(0)] - \sum_{k=1}^{M-1} 
  \frac{B_{2k}}{(2k)!}(\Delta t)^{2k}[f^{(2k-1)}(\Delta t) - f^{(2k-1)}(0)] 
  - \frac{(\Delta t)^{2M+1}}{(2M)!}B_{2M}f^{(2M)}(\eta\Delta t) \nonumber \\
  & = &   \frac{\Delta t}{2}[f(\Delta t)+f(0)] - \sum_{k=1}^{M-1} 
  \frac{B_{2k}}{(2k)!}(\Delta t)^{2k}[f^{(2k-1)}(\Delta t) - f^{(2k-1)}(0)]
  + {\cal O}((\Delta t)^{2M+1}),
\end{eqnarray}
where $0\leq\eta\leq 1$.
The $B_i, \ i =1,2,\dots$ are the Bernoulli numbers, i.e., $B_1=1/2,
B_2=1/6, B_3=0, B_4=-1/30, B_5=0, B_6=1/42, B_7=0, B_8=-1/30,\dots$.   It should be noted that if  $f(\theta)$ is not a polynomial, the Euler-MacLaurin formula  is  an asymptotic series~\cite[page 469]{jeffreys46}.

The first term of the sum includes the next
higher approximation compared to Eq.~(\ref{eq:1.5}).  We obtain
\begin{eqnarray}\label{eq:1.13}
    \displaystyle \psi(t+\Delta t) & = & e^{-iH\Delta t/\hbar}\psi(t) 
    -\frac{\textstyle i}{\textstyle \hbar}
    e^{-iH\Delta t/\hbar} \frac{\textstyle \Delta t}{\textstyle 2} 
    \left[e^{iH\Delta t/\hbar}
      N(t+\Delta t) + N(t)\right] \nonumber \\ 
    && \hspace{-0.4in}\displaystyle + \frac{i}{\hbar} e^{-iH\Delta t/\hbar}
    \frac{(\Delta t)^2}{12}\left[\frac{i}{\hbar}H e^{iH\Delta t/\hbar}
      N(t+\Delta t) +  e^{iH\Delta t/\hbar}N'(t+\Delta t) 
      - \frac{i}{\hbar}HN(t) - N'(t)\right] + {\cal O}((\Delta t)^5),
\end{eqnarray}
where the prime refers to differentiation with respect to $t$.
Rearranging the equation we get
\begin{eqnarray}\label{eq:1.15}
  \psi(t+\Delta t) && \ + \ \frac{i\Delta t}{2\hbar}N(t+\Delta t) 
  -\frac{i(\Delta t)^2}{12\hbar}\left[\frac{i}{\hbar}HN(t+\Delta t) 
    + N'(t+\Delta t)\right] \nonumber \\
  && = K_2^{(2)}K_1^{(2)}\left\{\psi(t) - \frac{i\Delta t}{2\hbar}N(t)
    -\frac{i(\Delta t)^2}{12\hbar}\left[\frac{i}{\hbar}HN(t)+N'(t)\right]
  \right\} + {\cal O}((\Delta t)^5),
\end{eqnarray}
\end{widetext}
where $K_s^{(M)}$ is defined in I as~[15] 
\begin{equation}\label{eq:1.16}
  K_s^{(M)} \equiv \frac{1+(\textstyle iH\Delta t/\hbar)/z_s^{(M)}} {1- (iH\Delta t/\hbar)/\bar{z}_s^{(M)}}.
\end{equation}
The order in which the operators $K_s^{(M)}$ are applied is not important
since they commute.  We define 
\begin{eqnarray}\label{eq:1.16a}
\Psi^{(+)} & \equiv & \Psi_{n+1}  = \psi_{n+1} + \dfrac{i\Delta t}{2\hbar}N_{n+1}
\nonumber \\ 
&& -\dfrac{i(\Delta t)^2}{12\hbar}\left[\dfrac{i}{\hbar}HN_{n+1} +
N'_{n+1}\right],
\end{eqnarray}
and 
\begin{equation}\label{eq:1.16b}
\Psi^{(-)}\equiv \Psi_n = \psi_n - \dfrac{i\Delta t}{2\hbar} N_n - \dfrac{i(\Delta t)^2}{12\hbar} \left[\dfrac{i}{\hbar}HN_n + N_n'\right].  
\end{equation}
Thus
\begin{equation}\label{eq:1.16c}
\Psi_{n+1} = K_2^{(2)}K_1^{(2)} \Psi_n.
\end{equation}
We use the known $\psi_n \approx \psi(x,t)$ to calculate $\Psi_n$ from Eq.~(\ref{eq:1.16b}).  Then we iteratively obtain $\Psi^{(+)}$ from $\Psi^{(-)}$ \`a la method described in I, i.e., $\Psi_{n+1/2} = K_1^{(2)}\Psi_n$ and $\Psi^{(+)} \equiv \Psi_{n+1} = K_2^{(2)}\Psi_{n+1/2}$.  From Eq.~(\ref{eq:1.16a}) we obtain $\psi_{n+1} \approx \psi(x,t+\Delta t)$.  The conversion from $\Psi^{(\pm)}$ to $\psi$ and vice versa occurs before and after the sequence of the iterative applications of the $K^{(M)}_s$ operators.

For known $N(x,t)$ Eq.~(\ref{eq:1.15}) can be solved in principle
using the method described in I.   Two new features  are the operation of $H$ on $N$ and the time differentiation
of $N(x,t)$.  The function $N(x,t)$ can be discretized in the same way
as $\psi(x,t)$ so that we form discrete elements $N_{n,j}\approx 
N(x_j,t_n)$.  In the discretized form
\begin{equation}\label{eq:1.17}
  (H\Psi_n)_j = -\frac{~~\hbar^2}{2m(\Delta x)^2}\sum_{k=-r}^r 
  c_k^{(r)}\psi_{n,j+k} + V_j\psi_{n,j}, 
\end{equation}
 where
\begin{equation}\label{eq:1.18}
  \begin{array}{lcl}
    e_k^{(r)} & = & -\frac{\textstyle ~~\hbar^2}
    {\textstyle 2m(\Delta x)^2}c_k^{(r)}, \\
    f_j^{(r)} & = & -\frac{\textstyle ~~\hbar^2}
    {\textstyle 2m(\Delta x)^2}c_0^{(r)} + V_j = e_0^{(r)} + V_j .
  \end{array}
\end{equation}
The coefficients $c_k^{(r)}$ are defined as in I.
The matrix form of $H$ is (suppressing the superscripts $^{(r)}$)
\begin{equation}
H = \left(\begin{array}{cccccccccc}
        f_0 & e_1 & e_2 & \cdots & e_r   & 0 \\
        e_1 & f_1 & e_1 &  \cdots & e_{r-1} & e_r \\
        e_2 & e_1 & f_2 & \cdots & e_{r-2} & e_{r-1} \\
        \vdots & \vdots & \vdots && \vdots & \vdots \\
        e_r & e_{r-1} & e_{r-2} & \cdots & f_r & e_1 \\
        0   & e_r & e_{r-1} & \cdots & e_1 & f_{r+1} \\
            &     &     &     &  && \ddots \\
            &     &     &     &  &&&       & f_{J-1} & e_1 \\
            &     &     &     &  &&&       & e_1     & f_J
        \end{array} \right).
\label{eq:1.19}
\end{equation}

The time partial derivative of $N(x,t)$ is straightforward if $N$ is
an analytically known function of $x$ and $t$.  If the function is
given in a discretized form, say $N_{n,j}$, accurate time derivatives may pose a challenge, especially the higher-order ones. 
\begin{widetext}

\subsection{Integration over time with arbitrary precision.}

 For the general case,  we start again with Eq.~(\ref{eq:1.4}),
\setcounter{equation}{4}
\begin{equation}\label{eq:1.4a}
  \psi(t+\Delta t) = e^{-iH\Delta t/\hbar}\psi(t) - \frac{i}{\hbar} 
  e^{-iH\Delta t/\hbar}
  \int_0^{\Delta t} e^{iH\theta/\hbar} N(t+\theta) \ d\theta.
\end{equation}
\setcounter{equation}{22}
Using the Euler-MacLaurin series~(\ref{eq:1.12}), we obtain
\begin{eqnarray}\label{eq:1.23}
  && \psi(t+\Delta t) \ = \ e^{-iH\Delta t/\hbar}\psi(t) - \frac{i\Delta t}{2\hbar}
  e^{-iH\Delta t/\hbar}\left[e^{iH\Delta t/\hbar}N(t+\Delta t) + N(t)\right]
  \nonumber \\
  &&\mbox{~~} +\frac{i}{\hbar}e^{-iH\Delta t/\hbar} 
  \sum_{k=1}^{M-1}\frac{B_{2k}}{(2k)!}(\Delta t)^{2k}\left\{
    \frac{\partial^{2k-1}~}{\partial\theta^{2k-1}}
   \left. \left.\left[e^{iH\theta/\hbar}N(t+\theta)\right]\right|_{\theta=\Delta t}
      - \frac{\partial^{2k-1}~}{\partial\theta^{2k-1}}\left[e^{iH\theta/\hbar}
      N(t+\theta)  \right]\right|_{\theta=0}\right\}
\end{eqnarray}
We note that $\theta$ is a time so that
$\left[H,\frac{\partial~}{\partial\theta}\right] = 0$.  We can
simplify the partial derivatives,
\begin{equation}\label{eq:1.24}
  \frac{\partial^{2k-1}~}{\partial\theta^{2k-1}} e^{iH\theta/\hbar}N(t+\theta)
  = e^{iH\theta/\hbar}\left(\frac{i}{\hbar}H + 
    \frac{\partial~}{\partial\theta}\right)^{2k-1}N(t+\theta).
\end{equation}
Using the binomial theorem, we obtain
\begin{equation}\label{eq:1.26}
  \left. \frac{\partial^{2k-1}~}{\partial\theta^{2k-1}} 
    e^{iH\theta/\hbar}N(t+\theta)
  \right|_{\theta=\Delta t}
  =e^{iH\Delta t/\hbar}\sum_{l=0}^{2k-1}\left(\begin{array}{c}
      2k-1 \\
      l
      \end{array}\right)
    \left(\frac{i}{\hbar}H\right)^{2k-1-l}
    \left. N^{(l)}(t+\Delta t)\right.,
\end{equation}
where $N^{(l)}$ is the $l$th partial derivative with respect to
$\theta$.  Similarly
\begin{equation}\label{eq:1.27}
   \left. \frac{\partial^{2k-1}~}{\partial\theta^{2k-1}} 
    e^{iH\theta/\hbar}N(t+\theta)
  \right|_{\theta=0}
  =\sum_{l=0}^{2k-1}\left(\begin{array}{c}
      2k-1 \\
      l
      \end{array}\right)
    \left(\frac{i}{\hbar}H\right)^{2k-1-l}
    N^{(l)}(t).
\end{equation}
Inserting the last two equations in Eq.~(\ref{eq:1.23}) we get
\begin{eqnarray}\label{eq:1.28}
  \psi(t+\Delta t) & = & e^{-iH\Delta t/\hbar}\psi(t) - \frac{i\Delta t}{2\hbar}
  e^{-iH\Delta t/\hbar}\left[e^{iH\Delta t/\hbar}N(t+\Delta t) + N(t)\right]
  \nonumber \\
  &&\mbox{~~} +\frac{i}{\hbar}e^{-iH\Delta t/\hbar} 
  \sum_{k=1}^{M-1}\frac{B_{2k}}{(2k)!}(\Delta t)^{2k}\left\{
    e^{iH\Delta t/\hbar} \sum_{l=0}^{2k-1} 
    \left(\begin{array}{c}
      2k-1 \\
      l
      \end{array}\right)
    \left(\frac{i}{\hbar}H\right)^{2k-1-l}
    N^{(l)}(t+\Delta t) \right. \nonumber \\
  &&\left. - \sum_{l=0}^{2k-1} \left(\begin{array}{c}
      2k-1 \\
      l
      \end{array}\right)
    \left(\frac{i}{\hbar}H\right)^{2k-1-l}
    N^{(l)}(t) \right\}.
\end{eqnarray}
We collect items evaluated at $t+\Delta t$ on the left side of the equation.
\begin{eqnarray}\label{eq:1.29}  
  && \psi(t+\Delta t) + \frac{i\Delta t}{2\hbar}N(t+\Delta t) - \frac{i}{\hbar}
  \sum_{k=1}^{M-1}\frac{B_{2k}}{(2k)!}(\Delta t)^{2k}
  \sum_{l=0}^{2k-1} 
    \left(\begin{array}{c}
      2k-1 \\
      l
      \end{array}\right)
    \left(\frac{i}{\hbar}H\right)^{2k-1-l}
    N^{(l)}(t+\Delta t) \nonumber \\
    && = e^{-iH\Delta t/\hbar} \left[\psi(t) -\frac{i\Delta t}{2\hbar}N(t)
        - \frac{i}{\hbar}
  \sum_{k=1}^{M-1}\frac{B_{2k}}{(2k)!}(\Delta t)^{2k}
  \sum_{l=0}^{2k-1} 
    \left(\begin{array}{c}
      2k-1 \\
      l
      \end{array}\right)
    \left(\frac{i}{\hbar}H\right)^{2k-1-l} N^{(l)}(t) \right]
\end{eqnarray}
We generalize the vector functions $\Psi^{(\pm)}_n$ by letting
\begin{equation}\label{eq:1.30}
  \Psi^{(\pm)}(x,t) = \psi(x,t) \pm \frac{i\Delta t}{2\hbar}N(x,t)
        - \frac{i}{\hbar}
  \sum_{k=1}^{M-1}\frac{B_{2k}}{(2k)!}(\Delta t)^{2k}
  \sum_{l=0}^{2k-1} 
    \left(\begin{array}{c}
      2k-1 \\
      l
      \end{array}\right)
    \left(\frac{i}{\hbar}H\right)^{2k-1-l} N^{(l)}(x,t) + {\cal O}(\Delta t^{2M+1}).
\end{equation}
\end{widetext}
With $M=2$, Eqs.~(\ref{eq:1.29}) and following are consistent with Eq.~(19) of Ref.~\cite{puzynin00} with error of ${\cal O}(\Delta t^5)$. 

We now express the time-evolution operator as (see Ref.~\cite{vandijk07})
\begin{equation}\label{eq:1.31}
  e^{-iH\Delta t/\hbar} = \prod_{s=1}^M K_s^{(M)} + {\cal O}((\Delta t)^{2M+1}).
\end{equation}
Since Eq.~(\ref{eq:1.29}) is equivalent to 
\begin{equation}\label{eq:1.31a}
\Psi^{(+)}\equiv \Psi_{n+1}  = e^{\textstyle -iH\Delta
  t/\hbar}\Psi_n,
\end{equation}
where $\Psi_n\equiv\Psi^{(-)}$,
 we write the relation as
\begin{equation}\label{eq:1.32}
  \Psi_{n+1} = \prod_{s=1}^M K_{s}^{(M)}\Psi_n.
\end{equation}
Defining  
\begin{equation}\label{eq:1.32a}
\Psi_{n+s/M} \equiv K_s^{(M)}\Psi_{n+(s-1)/M},
\end{equation}
we solve for $\Psi_{n+1}$ recursively, starting  from
\begin{equation}\label{eq:1:33}
 \Psi_{n+1/M} = K_1^{(M)}\Psi_n. 
\end{equation}  
Assuming that $\Psi^{(-)}\equiv\Psi_n$ is known from $\psi(x,t)$ and $N(x,t)$,
we determine $\Psi_{n+1/M}$ from Eq.~(\ref{eq:1:33}) which has a
form similar to that of Eq.~(\ref{eq:1.11}).  This is repeated to
obtain in succession
$\Psi_{n+2/M},\Psi_{n+3/M},\dots,\Psi_{n+(M-1)/M},\Psi_{n+1}\equiv
 \Psi^{(+)}$.
Since the operators $K_s^{(M)}$ commute, they can be applied in any
order.  Note that $\psi_{n+1,j}$ can be extracted from
$\Psi_{n+1}$ and $\Psi_n$ can be constructed from
$\psi_{n,j}$.  In each case it is assumed that $N(x,t)$ and its time
 derivatives are known.
The $\Psi^{(+)}_{n+1}$ is obtained from $\Psi^{(-)}_n$ by means
of a unitary operator.   Hence the normalization of the two functions is
the same, although 
this is in general not  so  for $\psi$.  Nevertheless the
integration process is stable.

\begin{figure}[h]
  \centering
   \resizebox{3.5in}{!}{\includegraphics{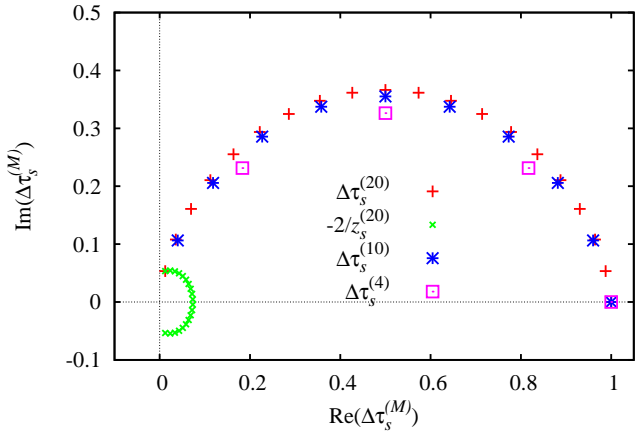}}
   \caption{(Color online) The $\Delta\tau_s^{(M)}$ for each $s=1,\dots,M$ from left to right plotted as dots on the complex plane. In this graph $M=20$, 10, and 4. The individual contributions for $M=20$, namely $-2/z_s^{(20)}$, are plotted as green dots. }
\label{fig:1}
\end{figure}

  Let us return to the time evolution within a step $\Delta t$,
\begin{equation}
\Psi_{n+s/M} = K_s^{(M)}\Psi_{n+(s-1)/M} =
 \prod_{s'=1}^sK_{s'}^{(M)}\Psi_n.  
\end{equation}
The form of the operator is~\footnote{There is an error in Ref.~\cite{vandijk07}. The plus and minus signs in Eq.~(3.4) of that paper should be interchanged.}
\begin{equation}
K_s^{(M)} = \dfrac{1+(iH\Delta t/\hbar)/z_s^{(M)}}{1-(iH\Delta t/\hbar)/\bar{z}_s^{(M)}}
\end{equation}
with $z_s^{(M)}$ a root of the numerator of the $[M/M]$ Pad\'e approximant of $e^z$.  In general $z_s^{(M)}$ is a complex number.  Nevertheless $K_s^{(M)}$ is a unitary operator.  In effect $K_s^{(M)}$ increases the time by a complex increment $-2\Delta t/z_s^{(M)}$.
As a check on the time increment formula one can show that  the roots $z_s^{(M)}$ for a particular $M$ obey the relationship
\begin{equation}
\sum_{s=1}^M \dfrac{1}{z_s^{(M)}} = -\dfrac{1}{2}.
\end{equation}
It is interesting to note that the times $t_n$, $t_{n+1}=t_n+\Delta t$, $t_{n+2}=t_n+2\Delta t$, etc., are real, but the intermediate times $t_n - 2\Delta t \sum_{s'=1}^s(1/z_{s'}^{(M)})$ are complex.  The times ``between" $t_n$ and $t_{n+1}$ can be denoted as
\begin{equation}
t_{n+s/M} = t_n - 2\Delta t \sum_{s'=1}^s\dfrac{1}{z_s^{(M)}}, \ \ \ s = 1,2,\dots, M.
\end{equation} 
We define the dimensionless time increment
\begin{equation}
\Delta \tau_s^{(M)} \equiv \dfrac{t_{n+s/M}-t_n}{\Delta t} = -2 \sum_{s'=1}^s\dfrac{1}{z_{s'}^{(M)}}. 
\end{equation}
 In Fig.~\ref{fig:1} we plot $\Delta\tau_s$ on the complex time plane.
The recursion~(\ref{eq:1.32a}) effectively gives us functions $\Psi_{n+s/M}$ which are related to the wave function at complex times; these wave functions need not be calculated since the iteration only involves the $\Psi_{n+s/M}$.  As we go through the $M$ operations of the $K_s^{(M)}$ we make, as it were, an excursion away from the real axis in the complex-time plane, but after the $M$th operation we are back on the real axis.  By placing the $z^{(M)}_s$ in different order we can follow different paths from the initial to final points; those paths however are not all as smooth.  One possible path is one which involves complex conjugates next to each other; then every other point lies on the real axis and points in between make excursions off the real axis.  However, since the operators $K^{(M)}_s$ commute the final point and time advance will be the same after completing a full 
time-step $\Delta t$.  The significance of this comment is that even when $N(x,t)$ is a real quantity with real arguments, in the calculation $N$ and $t$ need to be complex.  (See Eq.~(\ref{eq:1.30}).) 

As a final task we need to evaluate $N(x,t)$ and its partial time
derivatives.  Even if $N(x,t)$ is known analytically, only
for the simplest form can one write down the time
derivative of arbitrary order.  There may be problem specific-ways in which any-order time derivative can be obtained in a straightforward manner for more complex situations.

\subsection{Evaluation of the Bernoulli numbers}

The Bernoulli polynomials $B_n(x)$ are defined through the generating
function~\cite[formula 23.1.1]{abramowitz65}
\begin{equation}\label{eq:1.39}
  \frac{te^{xt}}{e^t -1} = \sum_{n=0}^{\infty} B_n(x)\frac{t^n}{n!}.
\end{equation}
The Bernoulli numbers are $B_n=B_n(0)$.  Some special values are
$B_0=1$, $B_1=-\frac{\textstyle 1}{\textstyle 2}$, and $B_{2n+1} = 0$
for $n = 1,2,\dots$.  The remaining Bernoulli numbers can calculated
using the Fourier expansion for the Bernoulli
polynomial~\cite[formula 23.1.16]{abramowitz65}
\begin{equation}\label{eq:1.40}
  B_n(x) = -2\frac{n!}{(2\pi)^n}\sum_{k=1}^\infty\frac{\cos(2\pi k x - 
\frac{1}{2}\pi n)}{k^n}, 
\end{equation}
which converges when $n>1, \ 0\leq x \leq 1$.  The Bernoulli numbers
occur when $x=0$ so that
\begin{equation}\label{eq:1.41}
  B_{2n} = -2(-1)^n \frac{(2n)!}{(2\pi)^{2n}}\sum_{k=1}^\infty\frac{1}
  {k^{2n}}.
\end{equation}
The following relations, due to Ramanujan, provide an efficient method for calculating Bernoulli numbers for even $m$: \\
for $ m\equiv 0\,\bmod\,6$,
\begin{equation}
     {{m+3}\choose{m}}B_m=\dfrac{m+3}{3}-\sum_{j=1}^{m/6}{m+3\choose{m-6j}}B_{m-6j}; 
\end{equation}
for $ m\equiv 2\,\bmod\,6$,
\begin{equation}
    {{m+3}\choose{m}}B_m=\dfrac{m+3}{3} -\sum_{j=1}^{(m-2)/6}{m+3\choose{m-6j}}B_{m-6j};
\end{equation}
and for $ m\equiv 4\,\bmod\,6$,
\begin{equation}
     {{m+3}\choose{m}}B_m=-\dfrac{m+3}{6} -\sum_{j=1}^{(m-4)/6}{m+3\choose{m-6j}}B_{m-6j}.
\end{equation}
As we observe from Table~\ref{table:3}, the Bernoulli numbers are increasing in magnitude with $n$.  This is a manifestation of the asymptotic nature of the Euler-MacLaurin series.  One expects the convergence of the series in Eq.~(\ref{eq:1.30}) to depend on the magnitude of $\Delta t$.
\begin{table}[h]
\begin{center}
\caption{The Bernoulli coefficients $B_n$.}
\label{table:3}
\begin{tabular}{cr|cr}
\hline\hline
~~~$n$~~~ & \multicolumn{1}{c|}{~~~~~~$B_n$~~~~~~} & ~~~$n$~~~ & \multicolumn{1}{c}{~~~~~~~~$B_n$~~~~~~} \\ \hline
0	&~~~~~~1.00000$\times 10^{+00}$ & 18	&5.49712$\times 10^{+01}$\\ 
2	&1.66667$\times 10^{-01}$ & 20	&-5.29124$\times 10^{+02}$\\
4	&-3.33333$\times 10^{-02}$& 22	&6.19212$\times 10^{+03}$\\
6	&2.38095$\times 10^{-02}$ & 24	&-8.65803$\times 10^{+04}$\\
8	&-3.33333$\times 10^{-02}$ & 26	&1.42552$\times 10^{+06}$\\
10	&7.57576$\times 10^{-02}$ & 28	&-2.72982$\times 10^{+07}$\\
12	&-2.53114$\times 10^{-01}$ & 30	&6.01581$\times 10^{+08}$\\
14	&1.16667$\times 10^{+00}$ & 32	&-1.51163$\times 10^{+10}$\\
16	&-7.09216$\times 10^{+00}$ & 34	&4.29615$\times 10^{+11}$\\ \hline
\end{tabular}
\end{center}
\end{table}

\subsection{Errors}

 In Ref.~\cite{vandijk07} we analyze truncation errors in the solution wave function obtained from the homogeneous equations.  These are expressed as
\begin{equation}\label{eq:101}
e^{(r)} = C^{(r)}(\Delta x)^{2r} \ \ \ \mathrm{and} \ \ \ e^{(M)} = C^{(M)}(\Delta t)^{2M+1},
\end{equation}
for the spatial and temporal dependencies.   The constants $C^{(r)}$ and $C^{(M)}$ are expected to be slowly varying functions of $r$ and $M$, respectively.  The variables $x$ and $t$ are independent.  When a particular precision of the wave function has been achieved in one variable, we can increase the order of approximation for the other variable and will reach that precision, saturating the process; the results will continue to be identical regardless how much more the order of the second variable is increased.  This is shown in Refs.~\cite{shao09,vandijk11}.      

For the solutions of the nonhomogeneous equations we have taken the same orders of approximation for the wave function and for the Euler-MacLaurin expansion.  Depending on the particular equation, it may be more efficient to consider different orders.  For instance, if $N(x,t)$ is much slower varying function of $x$ and $t$ than $\psi(x,t)$, lower orders in the Euler-MacLaurin expansion may be appropriate.  In our examples we do not know that ahead of time, so we use the same orders.  We do emphasize however that the constants $C^{(M)}$ and $C^{(r)}$ depend on the higher-order partial derivatives of the wave function and the source term, and hence are model dependent.
    
When the exact solution is known the error of the numerical calculation at final time $t_1$ can be obtained using the formula 
\begin{equation}\label{eq:error}
(e_2)^2 = \int_{x_0}^{x_J} dx \ |\psi(x,t_1)-\psi_\mathrm{exact}(x,t_1)|^2.
\end{equation}
A small value of $e_2$ is indicative of near equality of both the modulus and the phase of $\psi$ and $\psi_\mathrm{exact}$.  For this integral, and other integrals such as the normalization, we use the 
formula
\begin{equation}
\int_{x_0}^{x_J} dx \ f(x) = \Delta x \sum_{j=0}^J f(x_j).
\end{equation}
Peters and Maley~\cite{peters68} have shown that this formula is an approximation to the integral to  ${\cal O}(\Delta x^{2r+1})$ provided one includes correction terms which involve $f(x_i)$ where $i=0,1,\dots,r$ and $i=J-r,J-r+1,\dots,J$.  Since the correction terms depend only on the wave function near the extreme ends of the spatial range, they do not contribute significantly in our examples since the wave function is (nearly) zero there.  

In cases for which the exact solution is not known we can estimate the error by comparing the results for $M$ and $r$ with those for $M+1$ and $r+1$.  To that end we define the quantity
\begin{equation}\label{eq:eta}
(\eta_{M,r})^2 = \int_{x_0}^{x_{J}} dx |\psi^{(M,r)}(x,t_1) - \psi^{(M+1,r+1)}(x,t_1)|^2.
\end{equation}
Here the exact solution in Eq.~(\ref{eq:error}) is approximated by $\psi^{(M+1,r+1)}(x,t)$.   

\section{Numerical studies}
\label{sec:3}
\setcounter{equation}{0}
\subsection{Example 1: Non-spreading wave packet}

The examples for the numerical studies are chosen so that they have exact analytic solutions to which the numerical solutions can be compared.  They do not correspond in detail to actual physical systems.  Hopefully once the numerical method is validated, the method can be used for realistic systems.

Non-spreading or non-dispersive wave packets have been discussed and observed recently~\cite{stutzle05}.  Such ``Michelangelo" packets rely on an absorption process that removes the unwanted spreading part of the wave function so that  the packet retains its width and shape in coordinate space.  Earlier  non-spreading wave packets in free space, that are expressed in terms of Airy functions, were discussed by Berry and Balazs~\cite{berry79,besieris94}.  Somewhat related are the diffraction-free beams of particles for which there is no spreading in the transverse direction~\cite{durnin87,ryu14}. 

Given the results of our calculations, the stationary nonspreading wave packet provides as rigorous a test for the method as the travelling free (spreading) wave packet.  Thus in order \textit{to test the numerical procedure} we consider the stationary nonspreading wave packet,
\begin{equation}\label{eq:1.35}
\begin{split}
  \phi(x,t) & = (2\pi\sigma^2)^{-1/4} \\
&\times \exp\left[-\dfrac{(x-x_\mathrm{init})^2}{(2\sigma)^2}+ik_0(x-x_\mathrm{init})-\dfrac{i\hbar}{2m}k_0^2t\right],
\end{split}
\end{equation}
where $x_\mathrm{init}$ is the expectation value of the position of the wave
packet at time zero.  This wave packet is a solution of
\begin{equation}\label{eq:1.36}
  \left(i\hbar\frac{\partial~}{\partial t} +
	\frac{\hbar^2}{2m}\frac{\partial^2~}{\partial x^2}\right)
	\phi(x,t) = N(x,t)
\end{equation}
when
{\small
\begin{equation}\label{eq:1.37}
  N(x,t) = \frac{\hbar^2\left\{(x-x_\mathrm{init})^2 - 2\sigma^2
      [1+2ik_0(x-x_\mathrm{init})]\right\}}{8m\sigma^4}\phi(x,t).
\end{equation}
}
This nonspreading wave packet is also a solution of the time-dependent
homogeneous Schr\"odinger equation with the potential function
\begin{equation}\label{eq:1.38}
  V(x) = \frac{\hbar^2}{8m\sigma^4}\left\{(x-x_\mathrm{init})^2 - 2\sigma^2
      [1+2ik_0(x-x_\mathrm{init})]\right\}.
\end{equation} 

In the  example we used the same parameters as for the free travelling wave packet studied in I, and earlier in Ref.~\cite{goldberg67}, i.e.,  $\sigma=1/20$, $k_0=50\pi$, $x_0 = -0.5$, $x_J=1.5$, and $x_\mathrm{init} = 0.25$ with the units chosen so that $\hbar=2m=1$.  We set $\Delta t=2(\Delta x)^2$ and allow as much time as would be required for the free travelling packet to move from $x_\mathrm{init}=0.25$ to around 0.75.  In our case the packet does not move at all, but that does not detract from the validity of the test, since inaccurate calculations show definite movement of the packet.  
The results are tabulated in Table~\ref{table:1}.  The CPU time is  
the approximate time of computation and depends on the computer.  For the same computer the CPU times indicate relative times of computation.  We used two different computers, labelled as processor A (default, double precision) or B.  Times for different computers should not be compared.
\begin{table}[h]
\begin{center}
\caption{Summary of computational parameters and errors for example 1 with $k_0=50\pi$.  The quantity $\tau$ is the CPU time (processor A) in seconds and $\nu = (\Delta x)^2$.}
\label{table:1}
\begin{tabular}{ccccccrc}
\hline\hline
~$M$~~	&~~$r$~~	&~~$J$~~ &~~$\Delta t$	&$e_2$	& $\eta_{M,r}$ & $\tau$	&\\ \hline
1	&1	&2000	& ~$2\nu$ &$5.83\times 10^{-2}$~ & ~$5.82\times 10^{-2}$~	 &4.5	&\\
2	&2	&2000	&&$1.76\times 10^{-4}$ & $1.75\times 10^{-4}$	& 15	&\\
3	&3	&2000	&&$7.10\times 10^{-7}$ & $7.06\times 10^{-7}$	& 44 	&\\
4	&4	&2000	&&$3.19\times 10^{-9}$ & $3.18\times 10^{-9}$	& 110	&\\
5	&5	&2000	&&$1.59\times 10^{-11}$ &  $1.54\times 10^{-11}$	& 238	&\\
6	&6	&2000	&&$5.66\times 10^{-13}$ &  $9.35\times 10^{-14}$	& 457	&\\ \hline
2 & 19 & 260 & $2\nu$ & $1.98\times 10^{0}$ & $1.95\times 10^{0}$ & .20 \\
4	&19	&260 & & $5.13\times 10^{-3}$ &$4.99\times 10^{-3}$&.86	&\\
6	&19	&260 & & $2.90\times 10^{-6}$ &$2.85\times 10^{-6}$&2.5	&\\
8	&19	&260 & & $1.71\times 10^{-9}$ &$1.39\times 10^{-9}$&5.7	&\\
10 & 19 & 260 && $9.54\times 10^{-10}$ & $6.54\times 10^{-10}$ & .44K\footnotemark[1]\footnotetext[1]{Calculation done in quadruple precision with processor B.} \\
12 & 19 & 260 && $9.55\times 10^{-10}$ & $6.54\times 10^{-10}$ & .75K\footnotemark[1] \\
14 & 19 & 260 && $9.55\times 10^{-10}$ & $6.55\times 10^{-10}$ & 1.2K\footnotemark[1] \\
\hline
18&19 &260 & $\nu$ & $9.55\times 10^{-10}$ &$5.96\times 10^{-10}$&134	&\\
\hline
\end{tabular}
\end{center}
\end{table}

\begin{figure}[h]
  \centering
   \resizebox{3.5in}{!}{\includegraphics[angle=-90]{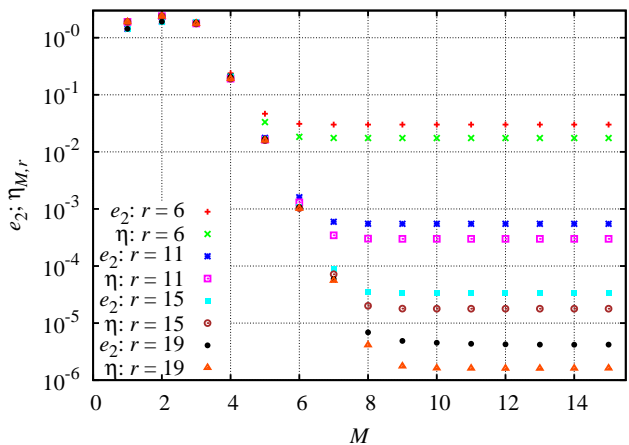}}
   \caption{(Color online) The errors (exact and estimated) for the calculations with the parameters of Table~\ref{table:1} including $\Delta t = 2\nu$ and $J=200$.  The $\eta$ in the legend of the graph refers to $\eta_{M,r}$.}
\label{fig:2}
\end{figure}

We graph the errors as a function of $M$ for given values of $r$ in Fig.~\ref{fig:2}.  For this graph we use the parameters of Table~\ref{table:1} with $2\nu$ and $J=200$.   The plateaus in the graph indicate a convergence of the error to a limit value.  The estimated error in this region tends to be smaller than the exact error.  However the two are of the same order of magnitude and, for the cases shown, the estimated error is no smaller than one-third the exact error.  This graph also indicates an approach to estimating the error when the exact solution is not known.  If one is in the region of the plateaus, increasing $M$ will not change the error, but increasing $r$ will move one to a lower plateau.  In a subsequent section we see similar plateaus for constant $M$ as $r$ is varied.  So to estimate the error by increasing both $M$ \textit{and} $r$ covers both instances. 

 We also considered a case with $k_0 = 1$, since that involves a smaller kinetic energy and a smaller (more reasonable) time derivative of $N(x,t)$.  The results are given in Table~\ref{table:2}.
\begin{table}[t]
\begin{center}
\caption{Same calculation as of Table~\ref{table:1} with $k_0=1$.}
\label{table:2}
\begin{tabular}{cccccccc}
\hline\hline
~$M$~~	&~~$r$~~	&~~$J$~~ &	$\Delta t$ &$e_2$	& $\eta_{M,r}$ & $\tau$	&\\ \hline
1	&1	&~~2000~~	& $2\nu$ & $1.25\times 10^{-5}$ & $1.25\times 10^{-5}$	&4.5\\
2	&2	&2000	&&$1.53\times 10^{-9}$ & $1.53\times 10^{-9}$	&15\\
3	&3	&2000	&&$3.04\times 10^{-13}$ & $3.04\times 10^{-13}$	&2.3K\footnotemark[1]\footnotetext[1]{Calculation done in quadruple precision with processor B}\\ \hline
4 & 19 & 260 & $2\nu$ & $1.90\times 10^{-13}$ & $1.90\times 10^{-13}$ & 43\footnotemark[1] \\
6 & 19 & 260 &  & $6.38\times 10^{-18}$ & $6.38\times 10^{-18}$ & .11K\footnotemark[1] \\
8 & 19 & 260 &  & $1.01\times 10^{-21}$ & $1.89\times 10^{-21}$ & .24K\footnotemark[1] \\
10 & 19 & 260 & & $1.16\times 10^{-20}$ & $6.39\times 10^{-20}$ & .44K\footnotemark[1] \\
\hline
\end{tabular}
\end{center}
\end{table}
  We expect the same qualitative behaviour for smaller values of $k_0$.  Such values of $k_0$ may make the differential equation less stiff and thus provide precise results with less effort. 
By judicious choice of the time and space discretization and orders of approximation one can obtain extremely accurate results.  For this example one needs values of $M>4$, whereas the traditional CN approach corresponds to $M=1$.
We note, however, that to obtain good results when $M\gtrsim 8$ the calculation need to be done in quadruple precision indicating that a substantial loss of significant figures in the computation occurs.  The likely reason for this is the need to raise the Hamiltonian matrix to  higher powers.  
\FloatBarrier

\subsection{Example 2: Coherent oscillations}

In this example we construct a nonhomogeneous Schr\"odinger equation from the one-dimensional harmonic oscillator,
\begin{equation}\label{eq:1.55}
-\dfrac{\hbar^2}{2m}\dfrac{\partial^2}{\partial x^2}\phi_\mathrm{nh}(x,t)+
\dfrac{1}{2}Kx^2\phi_\mathrm{nh}(x,t)=i\hbar\dfrac{\partial}{\partial t}\phi_\mathrm{nh}(x,t).
\end{equation}
The coherent oscillating wavepacket, that is an exact solution, is
\begin{eqnarray}\label{eq:1.56}
\phi_\mathrm{nh}(x,t) & = &\dfrac{\alpha^{1/2}}{\pi^{1/4}}\exp\left[-\dfrac{1}{2}(\xi-\xi_0\cos(\omega t))^2 \right. \nonumber \\
&& \hspace{-0.5in}\left. -i\left(\dfrac{1}{2}\omega t + \xi\xi_0\sin(\omega t) -\dfrac{1}{4}\xi_0^2\sin(2\omega t)\right)\right],
\end{eqnarray}
where $\omega=\sqrt{K/m}$, $\alpha = (mK/\hbar^2)^{1/4}$, $\xi=\alpha x$, and $\xi_0=\alpha a$.  The quantity $a$ is the initial position of the wave packet.
We can also consider $\phi_\mathrm{nh}(x,t)$ to be a particular solution of the nonhomogeneous Schr\"odinger equation
\begin{equation}\label{eq:1.57}
\left[i\hbar\dfrac{\partial}{\partial t}+\dfrac{\hbar^2}{2m}\dfrac{\partial^2}{\partial x^2}\right]\phi(x,t) = N(x,t)
\end{equation}
where
\begin{equation}\label{eq:1.58}
N(x,t)=\dfrac{1}{2}Kx^2\phi_\mathrm{nh}(x,t).
\end{equation}
The general solution of Eq.~(\ref{eq:1.57}) is the general solution of the associated homogeneous equation plus a particular solution of the nonhomogeneous equation.  The associated homogeneous equation is the free-particle equation which has as solution the free-particle wavepacket
\begin{eqnarray}
\phi_\mathrm{h}(x,t) & = & (2\pi\sigma^2)^{-1/4}[1+i\hbar t/(2m\sigma^2)]^{-1/2}\times  \nonumber \\ && \hspace{-0.7in}
\exp\left\{\dfrac{-(x-x_\mathrm{init})^2/(2\sigma)^2 + ik_0(x-x_\mathrm{init}) -i\hbar k_0^2t/(2m)}{1+i\hbar t/(2m\sigma^2)}\right\}. \nonumber \\ && 
\end{eqnarray}
 Thus a solution of Eq.~(\ref{eq:1.57}) is the superposition of the traveling free wavepacket and oscillating coherent wavepacket, i.e.,
\begin{equation}\label{eq:1.59}
\phi(x,t) = \phi_\mathrm{h}(x,t) + \phi_\mathrm{nh}(x,t).
\end{equation}  

An important consideration is the fact that we need to calculate the partial time derivative of various orders of the function $N(x,t)$.  Whereas in principle function (\ref{eq:1.58}) can be differentiated in closed form with respect to time an arbitrary number of times, such repeated differentiation
is not practical because of the complexity of the dependence of $\phi_\mathrm{nh}$ as a function of time.  Numerical differentiation becomes inaccurate quickly as the order increases.  However, $\phi_\mathrm{nh}(x,t)$ satisfies Eq.~(\ref{eq:1.55}) which we can write as
\begin{equation}\label{eq:1.60}
H_\mathrm{nh}\phi_\mathrm{nh}(x,t)=i\hbar\dfrac{\partial}{\partial t}
\phi_\mathrm{nh}(x,t). 
\end{equation}
Thus we obtain the $l$th time derivative of $N$ as
\begin{eqnarray}
\dfrac{\partial^l}{\partial t^l}N(x,t) & = & \dfrac{1}{2}Kx^2
\dfrac{\partial^l}{\partial t^l}\phi_\mathrm{nh}(x,t) \nonumber \\
& = & \dfrac{1}{2}Kx^2 
\left(\dfrac{-i}{\hbar}\right)^l H^l_\mathrm{nh}\phi_\mathrm{nh}(x,t).
\end{eqnarray}
The $l$th partial derivative obtained in this way is quite accurate and is
 (can be) obtained from the numerical wave function.

\begin{figure}[h]
\begin{flushleft}
\begin{list}{(\alph{bean})}{\usecounter{bean}\setlength{\rightmargin}{\leftmargin}\flushleft}
\item \centering
   \subfigure[
   ]{\resizebox{3.5in}{!}{\includegraphics{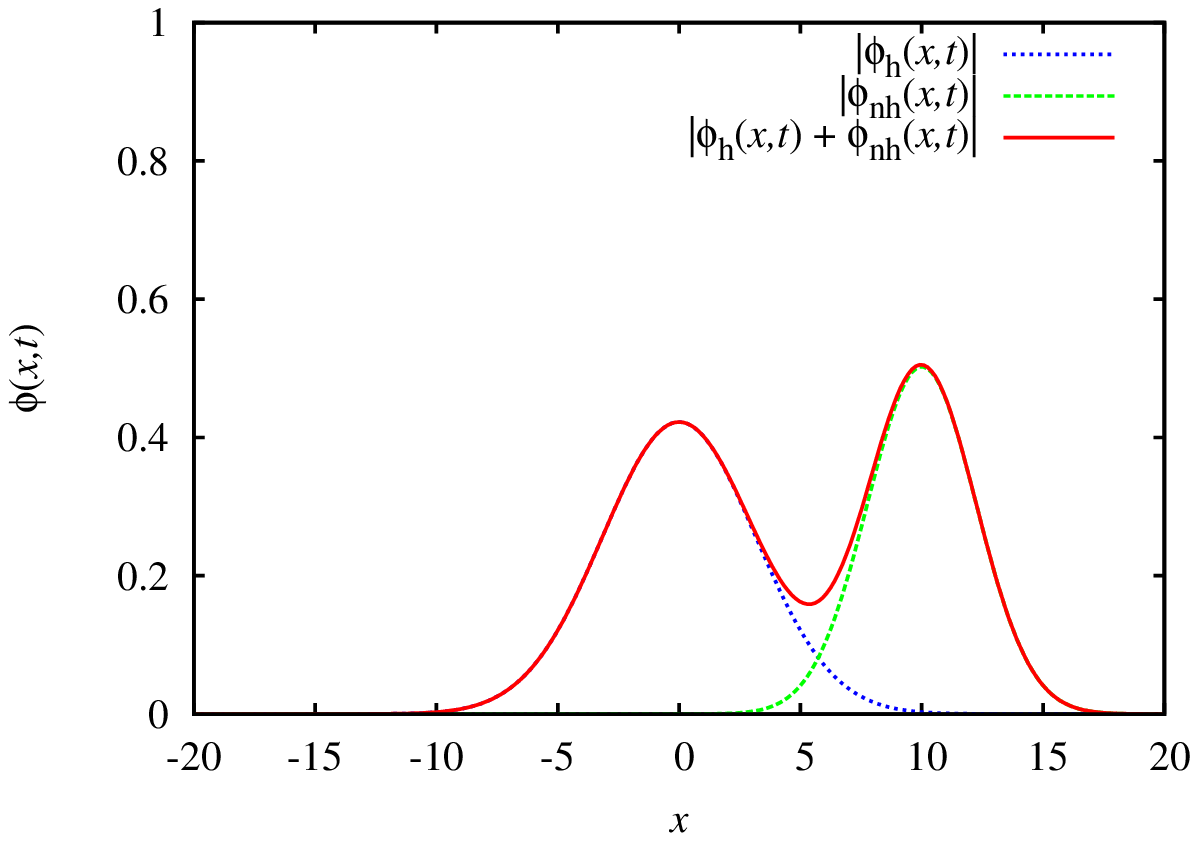}}}
\item \centering
   \subfigure[
   ]{\resizebox{3.5in}{!}{\includegraphics{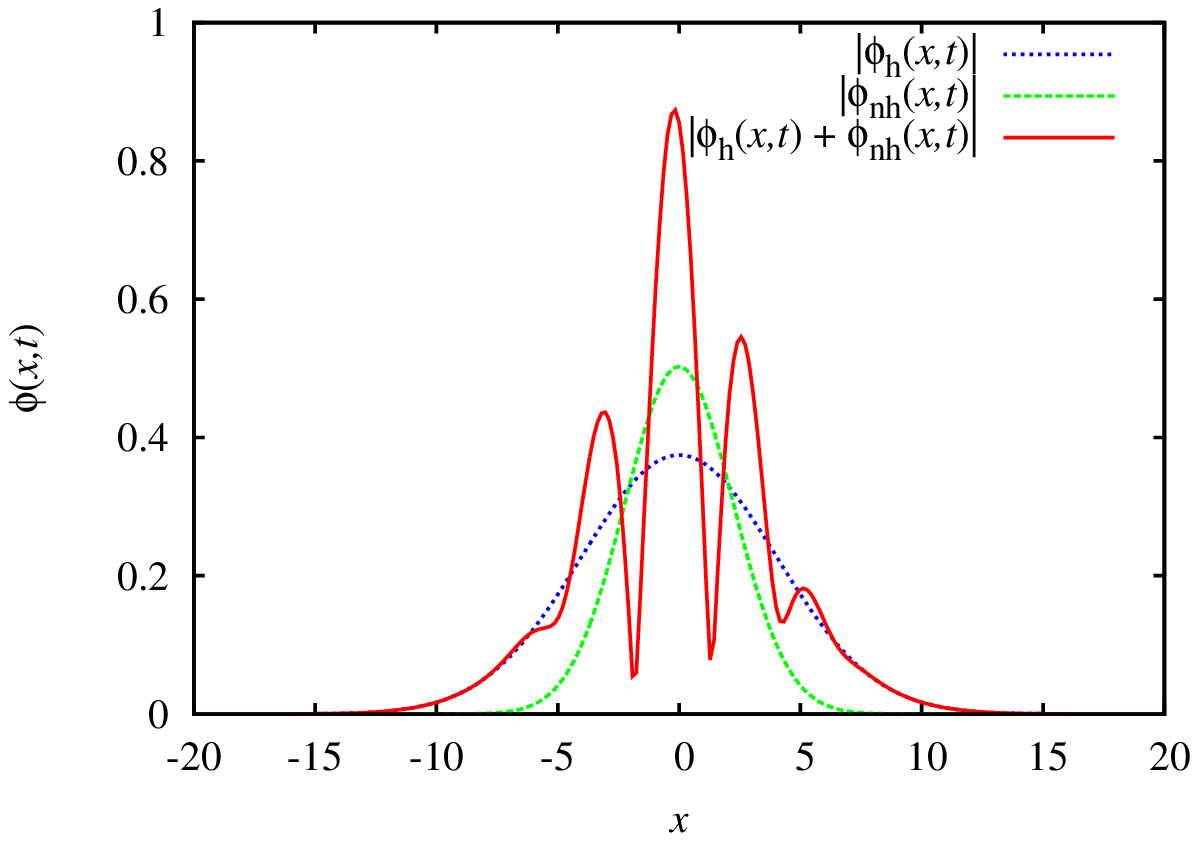}}}
   \caption{(Color online) The wave functions of example 2 at $t=0$ (panel (a)) and at $t=2.5\pi$ (panel (b)). The units used are such that $\hbar=m=1$.}
   \label{fig:3}
\end{list}
\end{flushleft}
\end{figure}

For the initial test we choose the parameters of I: $\omega=0.2$, $a=10$, $x_0=-80$, $x_J=80$ and final time $t_1=10\pi$, where the units are chosen so that $\hbar=m=1$.  From these parameters we determine $K$ and $\alpha$.  For the free wavepacket we choose $\sigma=1/\alpha$, 
 $k_0=0$ and $x_\mathrm{init} = 0$.  Thus the free wavepacket part of the wave function is stationary but is dispersing.  The interference of the two wavepackets can create significant oscillations in the overall wave function.  Figure~\ref{fig:3} shows the components and the total wave function at $t=0$ and at $t=2.5\pi$.
\newcounter{bean}
The results of the comparison of the exact and the numerical solutions are displayed in Table~\ref{table:4}.
\begin{table}[h]
\caption{The computational parameters and errors for example 2. The common parameters are  $\omega=0.2$, $a=10$, $x_0=-80$, $x_J=80$, $t_1=10\pi$, $dt=\pi/20$, $\sigma= 1/\alpha$, $k_0=0$ and $x_\mathrm{init} = 0$.  The units used are such that $\hbar=m=1$.
}
\label{table:4}
\begin{center}
\begin{tabular}{ccccccc}\hline\hline
$M$	&$r$	&$J$	&\multicolumn{1}{c}{$e_2$} & $\eta_{M,r}$	& $\tau$&\\ \hline
1	&1	&~~8000~~	&~~$1.67\times 10^{-1}$ &	$2.04\times 10^{-1}$&~~~~5.7	&\\
2	&2	&4000	&$7.21\times 10^{-4}$	& $8.78\times 10^{-4}$  &10	&\\
2	&2	&2000	&$8.54\times 10^{-4}$ &  $1.04\times 10^{-3}$	&3.7	& \\
2	&2	&1000	&$3.08\times 10^{-3}$ & $3.69\times 10^{-3}$	&1.7	&\\
4	&4	&1000	&$1.79\times 10^{-6}$ & $2.17\times 10^{-6}$	&14	&\\
6	&6	&1000	&$2.34\times 10^{-9}$ & $2.74\times 10^{-9}$	&58	&\\
8 & 8 & 1000 & $4.40\times 10^{-12}$ & $5.11\times 10^{-12}$ & 6.9K\footnotemark[1]\footnotetext[1]{Calculation done in quadruple precision with a different CPU, i.e., processor B.} \\
10 & 10 & 1000 & $3.33\times 10^{-14}$ & $1.26\times 10^{-14}$ & 16K\footnotemark[1] \\ 
12 & 12 & 1000 & $3.11\times 10^{-14}$ & $4.30\times 10^{-17}$ & 32K\footnotemark[1] \\
10 & 10& 800 & $8.57\times 10^{-13}$ & $9.59\times 10^{-13}$ & 12k\footnotemark[1] \\ \hline
10 & 10 & 300 & $5.69\times 10^{-5}$ & $3.80\times 10^{-5}$ & 110 \\
15 & 15 & 300 & $1.59\times 10^{-6}$ & $9.17\times 10^{-7}$ & 560 \\
19 & 19 & 300 & $1.44\times 10^{-7}$ & $7.54\times 10^{-8}$ & 1400 \\
\hline
\end{tabular}
\end{center}
\end{table} 

\section{Time-dependent Hamiltonian}  
\label{sec:4}
\setcounter{equation}{0}

In this section we examine time-dependent Hamiltonians, or rather time-dependent potential functions.   
 The method of I does not apply since it was assumed that the operators $K^{(M)}_s$ for different values of $s$ commute.  That is no longer the case  
if $H=H(t)$ is a function of time.  A more fundamental way of seeing that is that $H$ in the time-evolution operator is a function of time.  Successive operations of the evolution operators at different times introduce nonzero commutators of the Hamiltonian at different times.  In Ref.~\cite{puzynin00} the authors suggest an approach based on the Magnus expansion (see Ref.~\cite{blanes09} for a review), where one can in principle systematically obtain solutions with an error to ${\cal O}((\Delta t)^{2M})$, but beyond $M=2$ the method becomes cumbersome.  (See also Ref.~\cite{kormann08}.)  Our attempt involves considering the time-dependent potential term as the nonhomogeneous term in the equation, and then extract the wave function at the end of the time step from $\Psi^{(+)}$ by iteration.

We proceed as follows.  Suppose the equation to be solved is
\begin{equation}\label{eq:1.63}
\left[i\hbar\dfrac{\partial}{\partial t}-H_0-V(x,t)\right]\psi(x,t)=0,
\end{equation}
where $H_0$ could include another potential term which is independent of $t$.  We can rewrite the equation as
\begin{equation}\label{eq:1.64}
\left(i\hbar\dfrac{\partial}{\partial t}-H_0\right)\psi(x,t) = V(x,t)\psi(x,t)
\equiv N(x,t;\psi(x,t)).
\end{equation}
We explicitly indicate the dependence of $N$ on $\psi$.
At the beginning of a time interval $\psi(x,t)$ is known and we construct $\Psi^{(-)}(x,t)$ using Eq.~(\ref{eq:1.30}).  In the process we need the time derivatives of $N(x,t;\psi(x,t))=V(x,t)\psi(x,t)$.  We obtain them from
\begin{equation}\label{eq:1.65}
\dfrac{\partial^l}{\partial t^l}N(x,t;\psi(x,t)) = \sum_{l'=0}^l  
{l \choose l'} 
\dfrac{\partial^{l-l'}V}{\partial t^{l-l'}}\dfrac{\partial^{l'}\psi}{\partial t^{l'}}.
\end{equation}
The partial time derivatives of $V(x,t)$ need to be calculated analytically, but those of $\psi$ can be obtained using the following approach.  We rewrite Eq.~(\ref{eq:1.63})
\begin{equation}\label{eq:1.65.1}
\dfrac{\partial~}{\partial t}\psi(x,t) = A(x,t)\psi(x,t),
\end{equation}
where $A(x,t) = \left(-\dfrac{i}{\hbar}\right)[H_0(x)+V(x,t)]$.  We form a recursion to obtain the $l$th partial derivative with respect to $t$, i.e.,
\begin{equation}\label{eq:1.65.2}
\dfrac{\partial^l~}{\partial t^l}\psi(x,t) = f_l(A)\psi(x,t)
\end{equation} 
with
\begin{equation}\label{eq:1.66}
f_0=1 \ \ \ \mathrm{and} \ \ \   f_l(A) = \dfrac{\partial f_{l-1}}{\partial t} + f_{l-1}(A)A, 
\end{equation}
for $l=1,2,3,\dots$.
The first five function $f_l(A)$ are
\begin{equation}\label{eq:1.66a}
\begin{split}
f_1(A) & = A \\
f_2(A) & = A^2 + \dfrac{\partial A}{\partial t} \\
f_3(A) & = A^3 + A\dfrac{\partial A}{\partial t}  + 2\dfrac{\partial A}{\partial t} A + \dfrac{\partial^2 A}{\partial t^2} \\
f_4(A) & = A^4 + A^2\dfrac{\partial A}{\partial t} + 2A\dfrac{\partial A}{\partial t}A + 3\dfrac{\partial A}{\partial t}A^2 + A\dfrac{\partial^2 A}{\partial t^2} \\ 
&~~~~ +3\left(\dfrac{\partial A}{\partial t}\right)^2 + 3\dfrac{\partial^2 A}{\partial t^2}A + \dfrac{\partial^3 A}{\partial t^3}   \\
f_5(A) & = A^5 + A^3\dfrac{\partial A}{\partial t} + 2A^2\dfrac{\partial A}{\partial t} A + 3 A \dfrac{\partial A}{\partial t} A^2 + 4\dfrac{\partial A}{\partial t}A^3 \\ 
 &~~~~ + A^2 \dfrac{\partial^2 A}{\partial t^2} + 3A\dfrac{\partial^2 A}{\partial t^2}A  + 3A\left(\dfrac{\partial A}{\partial t}\right)^2 + 4\dfrac{\partial A}{\partial t} A \dfrac{\partial A}{\partial t}  \\
& ~~~~~ +8\left(\dfrac{\partial A}{\partial t}\right)^2A + 6\dfrac{\partial^2 A}{\partial t^2}A^2 + A\dfrac{\partial^3 A}{\partial t^3} + 4\dfrac{\partial A}{\partial t}\dfrac{\partial^2 A}{\partial t^2}  \\
& ~~~~~ + 6\dfrac{\partial^2 A}{\partial t^2}\dfrac{\partial A}{\partial t} + 4 \dfrac{\partial^3 A}{\partial t^3}A +\dfrac{\partial^4 A}{\partial t^4}
\end{split}
\end{equation} 
Any order of the derivative of the wave function can be obtained, but in practice the formulas become increasingly more onerous to work with as $l$ increases. 

After one time increment we obtain $\Psi^{(+)}(x,t)$ at the incremented time.  From it we extract the new $\psi(x,t)$.  (Note that $t\rightarrow t+\Delta t$, but for convenience we write $t$.)  Thus
\begin{equation}\label{eq:1.68}
\psi(x,t) 
=  \Psi^{(+)}(x,t) - \dfrac{i\Delta t}{2\hbar}N(x,t;\psi(x,t)) + F(x,t;\psi(x,t))
\end{equation}
where
\begin{eqnarray}\label{eq:1.69}
F(x,t;\psi(x,t)) & = & \sum_{k=1}^{M-1}\frac{B_{2k}}{2k}(\Delta t)^{2k} \times \nonumber \\
&&\hspace{-1.0in} 
  \sum_{l=0}^{2k-1} \frac{1}{(2k-1-l)!l!} 
    \left(\frac{i}{\hbar}\right)^{2k-l}H_0^{2k-1-l}
    N^{(l)}(x,t;\psi(x,t)) \nonumber \\
&&
\end{eqnarray}
for $M\geq 2$.
Note that $F(x,t;\psi(x,t))$ is at least ${\cal O}((\Delta t)^2)$.  Thus we can write
\begin{equation}\label{eq:1.70}
\psi(x,t) = \dfrac{\Psi^{(+)}(x,t) + F(x,t;\psi(x,t))}{1 + \dfrac{i\Delta t}
{2\hbar}V(x,t)}.
\end{equation}
We solve this equation iteratively by making an initial approximation
\begin{equation}\label{eq:1.71}
\psi^{(0)}(x,t) = \dfrac{\Psi^{(+)}(x,t) + F(x,t;\psi(x,t-\Delta t))}{1 + \dfrac{i\Delta t}
{2\hbar}V(x,t)},
\end{equation}
and evaluating successively
\begin{equation}\label{eq:1.72}
\psi^{(i+1)}(x,t) = \dfrac{\Psi^{(+)}(x,t) + F(x,t;\psi^{(i)}(x,t))}{1 +
\dfrac{i\Delta t}{2\hbar}V(x,t)}, \ i = 0, 1, \dots
\end{equation}
We continue the process until $e^{(i)}$ defined as
\begin{equation}
(e^{(i)})^2 = \int_{x_0}^{x_J} dx \ |\psi^{(i+1)}(x,t)-\psi^{(i)}(x,t)|^2
\end{equation}
is smaller than a prescribed amount.  In other words we look for the convergence
\begin{equation}\label{eq:1.74}
\lim_{i\rightarrow\infty} \psi^{(i)}(x,t) = \psi(x,t).
\end{equation}

\subsection{Example 3: Time-dependent oscillator}
As a last example we will consider the harmonic oscillator with time-dependent 
frequency~\cite{husimi53}. (See also Ref.~\cite{moyacessa03} and references contained in it.)    The potential has the form
\begin{equation}\label{eq:1.75}
V(x,t) = \dfrac{1}{2}m\omega^2(t)x^2
\end{equation}
with $\omega^2(t)=\omega_0^2(1-f e^{-\mu t})$ where $f$ is a positive proper fraction and $\mu$ a positive number.  Systems with such potentials are known to have analytic solutions and we will compare the numerical solution to the analytical one.  The time dependence of $V(x,t)$ is of such a nature that partial derivatives with respect to time can easily be obtained.  On the other hand by choosing different values of $\mu$ we can make the potential term vary slowly or rapidly with time.

\begin{figure}[h]
   \rotatebox{-90}{\resizebox{2.5in}{!}{\includegraphics{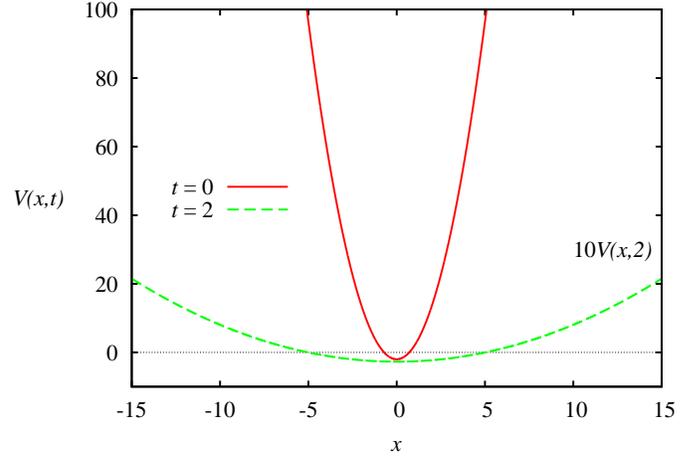}}}
   \caption{(Color online) The potential of Eq.~(\ref{eq:1.76}) as a function of $x$ at $t=0$ 
and $t=2$.  Since $V(x,2)$ is relative small, the curve plotted is magnified by a factor of ten.  The units used are such that $\hbar=2m=1$.}
\label{fig:4}
\end{figure}

\begin{figure}[h]
   \rotatebox{-90}{\resizebox{2.5in}{!}{\includegraphics{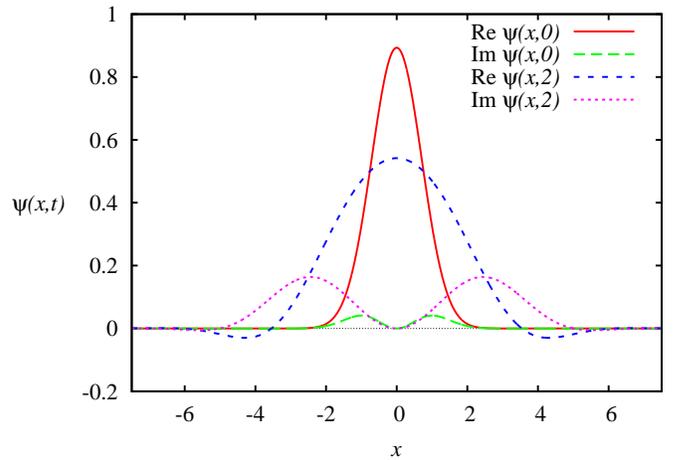}}}
   \caption{(Color online) The wave functions  Eq.~(\ref{eq:1.77}) as a function of $x$ at $t=0$ and $t=2$.   The units used are such that $\hbar=2m=1$. }
\label{fig:5}
\end{figure}

A simpler potential can be obtained using the method of Fityo and Tkachuk~\cite{fityo05}.  When $\hbar=1$ and $m=1/2$ the potential 
\begin{equation}\label{eq:1.76}
V(x,t) = \left(4e^{-2t} - \dfrac{1}{16}\right)x^2 - 2e^{-t}
\end{equation}
yields a normalized wave function
\begin{equation}\label{eq:1.77}
\psi(x,t) = \left(\dfrac{2}{\pi}\right)^{1/4}
\exp\left(-x^2e^{-t}-\dfrac{1}{4}t + \dfrac{i}{8}x^2\right).
\end{equation}
The advantage of this potential as a test case is that it is easily differentiable with respect to time to any order.  In order to obtain the derivates of $N(x,t)$ we use Eqs.~(\ref{eq:1.65}) and (\ref{eq:1.66}) with
\begin{equation}\label{eq:1.78}
\dfrac{\partial^lV}{\partial t^l} = (-1)^l(2^{l+2}e^{-2t}x^2 - 2e^{-t})
\end{equation}
for $l\geq 1$.
To obtain a feel for the potential and the wave function they are plotted in Figs.~\ref{fig:4} and \ref{fig:5} respectively.

The sample calculations were done with $-15\leq x \leq 15$ and a final time of $t=2$.  The initial wave function was that of Eq.~(\ref{eq:1.77}) with $t=0$.  The
results are tabulated in Table~\ref{table:5}. 
\begin{table}[h]
\begin{center}
\caption{Parameters and errors for example 3.}
\label{table:5}
\begin{tabular}{cccrcccc}
\hline\hline
$M$       &$r$      &$J$      &$\Delta t$        &$e_2$ &$\eta_{M,r}$  & $\tau$\footnotemark[1]\footnotetext{Time to calculate $e_2$ in quadruple precision with processor B.} &\\ \hline
1       &~19~      &200    &0.0075   &~~~$3.22035\times 10^{-5}$~~    & $3.22173\times 10^{-5}$   & 27& \\
        &       &       &0.0010  &$5.72355\times 10^{-7}$ &  $5.72353\times 10^{-7}$  &210   & \\
        &       &       &~0.0001 &$5.72356\times 10^{-9}$    &   & 1.9K & \\
2       &       &       &0.0075   &$7.60367\times 10^{-9}$    &  $7.60056\times 10^{-9}$  & 91 & \\
        &       &       &0.0010  &$2.40331\times 10^{-12}$  &$2.40328\times 10^{-12}$ & 590   &  \\
3       &       &       &0.0075   &$3.85317\times 10^{-12}$    &   $3.84974\times 10^{-12}$ & 310 & \\
        &       &       &0.0010	&$7.16318\times 10^{-15}$& $2.20551\times 10^{-17}$ 	&1.8K  & \\
4       &       &       &0.0075   &$8.77841\times 10^{-15}$   &   & 1.2K &  \\ 
        &       &       &0.0010	&$7.16561\times 10^{-15}$ &	&   6.5K &  \\ \hline
\end{tabular}
\end{center}
\end{table}

We observe an increase in efficiency and precision of the calculation with increasing values of $M$ for relatively small $M$.  When $M$ is larger than 3 or 4, increasing the size of $\Delta t$ gives unstable solutions in the sense that the
convergence of the fixed point iteration~(\ref{eq:1.72}) does not occur.  The criterion of convergence that we used is that the iterative procedure is terminated when $e^{(i)}<10^{-20}$.  

\begin{figure}[h]
  \centering
   \resizebox{3.5in}{!}{\includegraphics[angle=-90]{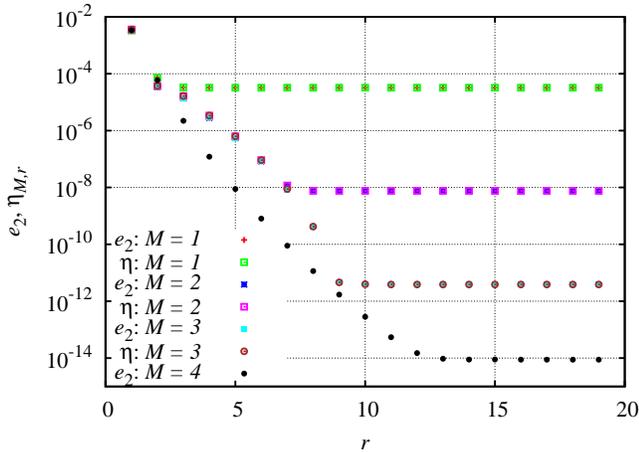}}
   \caption{(Color online) The errors (exact and estimated) for the calculations with the parameters $x_0=-15,x_J=15, J = 200, \Delta t=0.0075$.  The units used are such that $\hbar=1, m=1/2$.  For the $M=3$ case we terminate iteration for $i$ when $e^{(i)}<10^{-19}$.  The $\eta$ in the legend of the graph refers to $\eta_{M,r}$.}
\label{fig:6}
\end{figure}

In Fig.~\ref{fig:6} we show the deviations from the exact solutions $e_2$ and the deviations from the next higher order approximation $\eta_{M,r}$.  We obtain plateaus on which the values are very nearly the same.  For instance for $M=2$   for the six highest values of $r$ each of the two errors are identical to seven significant figures and $e_2=\eta_{M,r}$ to three significant figures.  On the graph all the values of $e_2$ and $\eta_{M,r}$ for the same parameters are indistinguishable.

(As an aside, another candidate to test the method for time-dependent potentials is the linear time-dependent potential discussed by Guedes~\cite{guedes01}.)

\section{Concluding remarks}
\label{sec:5}

We have developed accurate numerical methods for solving the TDSE  with sources and/or time-dependent potentials.  Since the function  that is evaluated numerically, i.e., $\Psi^{(+)}$, from which the solution is extracted, is a normalized function the generalized CN method provides for a stable procedure.  The examples of exactly solvable systems  indicate that extremely accurate numerical solutions can be obtained.  We employed double precision in the initial calculations, but with quadruple precision truncation errors could be driven down further (see Ref.~\cite{vandijk11}).  There is a caveat, however, since the Euler-MacLaurin series is an asymptotic series;  for a given $\Delta t$ increasing the order of the approximation will eventually cause the precision to decrease.  

The method allows for a calculation to arbitrary order of $\Delta t$ and of $\Delta x$.  In the calculations of the examples we calculate an error which corresponds to the deviation from the exact solution.  However, considering the difference of the numerical solution with one that is one order higher in both variables, we have an estimate of the error which is of the same order of magnitude as the actual error.  This permits one to estimate errors when no exact analytic solution is available.

The calculations were done in one dimension.  They would be similar for partial wave calculations as was done in I.  A natural extension is to consider systems of coupled equations as in Ref.~\cite{houfek09}.  Two or three dimensional calculations will be explored in future work.      

\acknowledgments
We thank Professors Y. Nogami and D.W.L. Sprung for helpful discussions and constructive comments on the manuscript.  We are grateful for support from the Natural Sciences and Engineering Council of Canada during the initial stages of this research.

\vspace{\baselineskip}

\appendix

\section{Harmonic oscillator with time-dependent frequency}
\renewcommand{\theequation}{\Alph{section}.\arabic{equation}}
To obtain potential (\ref{eq:1.76}) we use the method of Fityo and Tkachuk~\cite{fityo05},~\footnote{Equation~(16) of Ref.~\cite{fityo05} has a typo: the 3 should be replaced by 2.}. Using their notation, we choose 
\begin{equation}\label{eq:a.1}
\tilde{f}(x,t)=x^2e^{-t}.
\end{equation}
Thus we obtain
\begin{equation}\label{eq:a.2}
F(t) = \int_{-\infty}^\infty e^{-2\tilde{f}(x,t)} \ dx =\sqrt{\dfrac{\pi}{2}}e^{t/2},
\end{equation}
which leads to
\begin{equation}\label{eq:a.3}
f(x,t) = \tilde{f} + \dfrac{1}{2}\ln F = x^2e^{-t}+\dfrac{1}{4}t +\dfrac{1}{4}\ln{\pi/2}.
\end{equation}
We generate $g(x,t)$ from Eq.~{(5)} of Ref.~\cite{fityo05}
\begin{equation}\label{eq:a.4}
g(x,t) = \dfrac{1}{2}\int_0^x \ e^{2f(y,t)} \dfrac{\partial~}{\partial t}
\int_{-\infty}^y \ e^{-2f(z,t)} \ dzdy = -\dfrac{1}{8}x^2.
\end{equation}
Note that we remove a (constant) additive infinity from $g(x,t)$ by integrating over $y$ from $0$ to $x$ rather than from $-\infty$ to $x$ as in Ref.~\cite{fityo05}.  This has no consequence for the potential, but eliminates a (constant) infinite phase from the wave function.

The potential is obtained from
\begin{equation}\label{eq:a.5}
V(x,t) = g_t + f_x^2 - g_x^2 - f_{xx}
\end{equation}
and the normalized wave function is
\begin{equation}\label{eq:a.6}
\psi(x,t) = e^{-f(x,t)-ig(x,t)}.
\end{equation}
It is straightforward to verify that this wave function satisfies the time-dependent Schr\"odinger equation with potential~(\ref{eq:a.5}).


\begin{thebibliography}{33}%
\makeatletter
\providecommand \@ifxundefined [1]{%
 \@ifx{#1\undefined}
}%
\providecommand \@ifnum [1]{%
 \ifnum #1\expandafter \@firstoftwo
 \else \expandafter \@secondoftwo
 \fi
}%
\providecommand \@ifx [1]{%
 \ifx #1\expandafter \@firstoftwo
 \else \expandafter \@secondoftwo
 \fi
}%
\providecommand \natexlab [1]{#1}%
\providecommand \enquote  [1]{``#1''}%
\providecommand \bibnamefont  [1]{#1}%
\providecommand \bibfnamefont [1]{#1}%
\providecommand \citenamefont [1]{#1}%
\providecommand \href@noop [0]{\@secondoftwo}%
\providecommand \href [0]{\begingroup \@sanitize@url \@href}%
\providecommand \@href[1]{\@@startlink{#1}\@@href}%
\providecommand \@@href[1]{\endgroup#1\@@endlink}%
\providecommand \@sanitize@url [0]{\catcode `\\12\catcode `\$12\catcode
  `\&12\catcode `\#12\catcode `\^12\catcode `\_12\catcode `\%12\relax}%
\providecommand \@@startlink[1]{}%
\providecommand \@@endlink[0]{}%
\providecommand \url  [0]{\begingroup\@sanitize@url \@url }%
\providecommand \@url [1]{\endgroup\@href {#1}{\urlprefix }}%
\providecommand \urlprefix  [0]{URL }%
\providecommand \Eprint [0]{\href }%
\providecommand \doibase [0]{http://dx.doi.org/}%
\providecommand \selectlanguage [0]{\@gobble}%
\providecommand \bibinfo  [0]{\@secondoftwo}%
\providecommand \bibfield  [0]{\@secondoftwo}%
\providecommand \translation [1]{[#1]}%
\providecommand \BibitemOpen [0]{}%
\providecommand \bibitemStop [0]{}%
\providecommand \bibitemNoStop [0]{.\EOS\space}%
\providecommand \EOS [0]{\spacefactor3000\relax}%
\providecommand \BibitemShut  [1]{\csname bibitem#1\endcsname}%
\let\auto@bib@innerbib\@empty
\bibitem [{\citenamefont {van Dijk}\ and\ \citenamefont
  {Toyama}(2007)}]{vandijk07}%
  \BibitemOpen
  \bibfield  {author} {\bibinfo {author} {\bibfnamefont {W.}~\bibnamefont {van
  Dijk}}\ and\ \bibinfo {author} {\bibfnamefont {F.~M.}\ \bibnamefont
  {Toyama}},\ }\href@noop {} {\bibfield  {journal} {\bibinfo  {journal} {Phys.
  Rev. E}\ }\textbf {\bibinfo {volume} {75}},\ \bibinfo {pages} {036707}
  (\bibinfo {year} {2007})}\BibitemShut {NoStop}%
\bibitem [{\citenamefont {Wang}\ and\ \citenamefont {Shao}(2009)}]{wang09}%
  \BibitemOpen
  \bibfield  {author} {\bibinfo {author} {\bibfnamefont {Z.}~\bibnamefont
  {Wang}}\ and\ \bibinfo {author} {\bibfnamefont {H.}~\bibnamefont {Shao}},\
  }\href@noop {} {\bibfield  {journal} {\bibinfo  {journal} {Comp. Phys.
  Comm.}\ }\textbf {\bibinfo {volume} {180}},\ \bibinfo {pages} {842} (\bibinfo
  {year} {2009})}\BibitemShut {NoStop}%
\bibitem [{\citenamefont {Tal-Ezer}\ and\ \citenamefont
  {Kosloff}(1984)}]{talezer84}%
  \BibitemOpen
  \bibfield  {author} {\bibinfo {author} {\bibfnamefont {H.}~\bibnamefont
  {Tal-Ezer}}\ and\ \bibinfo {author} {\bibfnamefont {R.}~\bibnamefont
  {Kosloff}},\ }\href@noop {} {\bibfield  {journal} {\bibinfo  {journal} {J.
  Chem. Phys.}\ }\textbf {\bibinfo {volume} {81}},\ \bibinfo {pages} {3967}
  (\bibinfo {year} {1984})}\BibitemShut {NoStop}%
\bibitem [{\citenamefont {van Dijk}\ \emph {et~al.}(2011)\citenamefont {van
  Dijk}, \citenamefont {Brown},\ and\ \citenamefont {Spyksma}}]{vandijk11}%
  \BibitemOpen
  \bibfield  {author} {\bibinfo {author} {\bibfnamefont {W.}~\bibnamefont {van
  Dijk}}, \bibinfo {author} {\bibfnamefont {J.}~\bibnamefont {Brown}}, \ and\
  \bibinfo {author} {\bibfnamefont {K.}~\bibnamefont {Spyksma}},\ }\href@noop
  {} {\bibfield  {journal} {\bibinfo  {journal} {Phys. Rev. E}\ }\textbf
  {\bibinfo {volume} {84}},\ \bibinfo {pages} {056703} (\bibinfo {year}
  {2011})}\BibitemShut {NoStop}%
\bibitem [{\citenamefont {Form\'anek}\ \emph {et~al.}(10)\citenamefont
  {Form\'anek}, \citenamefont {V\'a{\v n}a},\ and\ \citenamefont
  {Houfek}}]{formanek10}%
  \BibitemOpen
  \bibfield  {author} {\bibinfo {author} {\bibfnamefont {M.}~\bibnamefont
  {Form\'anek}}, \bibinfo {author} {\bibfnamefont {M.}~\bibnamefont {V\'a{\v
  n}a}}, \ and\ \bibinfo {author} {\bibfnamefont {K.}~\bibnamefont {Houfek}},\
  }in\ \href@noop {} {\emph {\bibinfo {booktitle} {Numerical Analysis and
  Applied Mathematics, International Conference 2010}}},\ \bibinfo {editor}
  {edited by\ \bibinfo {editor} {\bibfnamefont {T.~E.}\ \bibnamefont {Simos}},
  \bibinfo {editor} {\bibfnamefont {G.}~\bibnamefont {Psihoyios}}, \ and\
  \bibinfo {editor} {\bibfnamefont {C.}~\bibnamefont {Tsitouras}}}\ (\bibinfo
  {publisher} {American Institute of Physics},\ \bibinfo {year} {10})\ pp.\
  \bibinfo {pages} {667--670}\BibitemShut {NoStop}%
\bibitem [{\citenamefont {Ndong}\ \emph {et~al.}(2009)\citenamefont {Ndong},
  \citenamefont {{Tal-Ezer}}, \citenamefont {Kosloff},\ and\ \citenamefont
  {Koch}}]{ndong09}%
  \BibitemOpen
  \bibfield  {author} {\bibinfo {author} {\bibfnamefont {M.}~\bibnamefont
  {Ndong}}, \bibinfo {author} {\bibfnamefont {H.}~\bibnamefont {{Tal-Ezer}}},
  \bibinfo {author} {\bibfnamefont {R.}~\bibnamefont {Kosloff}}, \ and\
  \bibinfo {author} {\bibfnamefont {C.~P.}\ \bibnamefont {Koch}},\ }\href@noop
  {} {\bibfield  {journal} {\bibinfo  {journal} {J. Chem. Phys.}\ }\textbf
  {\bibinfo {volume} {130}},\ \bibinfo {pages} {124108} (\bibinfo {year}
  {2009})}\BibitemShut {NoStop}%
\bibitem [{\citenamefont {Ndong}\ \emph {et~al.}(2010)\citenamefont {Ndong},
  \citenamefont {Tal-Ezer}, \citenamefont {Kosloff},\ and\ \citenamefont
  {Koch}}]{ndong10}%
  \BibitemOpen
  \bibfield  {author} {\bibinfo {author} {\bibfnamefont {M.}~\bibnamefont
  {Ndong}}, \bibinfo {author} {\bibfnamefont {H.}~\bibnamefont {Tal-Ezer}},
  \bibinfo {author} {\bibfnamefont {R.}~\bibnamefont {Kosloff}}, \ and\
  \bibinfo {author} {\bibfnamefont {C.~P.}\ \bibnamefont {Koch}},\ }\href@noop
  {} {\bibfield  {journal} {\bibinfo  {journal} {J. Chem. Phys.}\ }\textbf
  {\bibinfo {volume} {132}},\ \bibinfo {pages} {064105} (\bibinfo {year}
  {2010})}\BibitemShut {NoStop}%
\bibitem [{\citenamefont {{van Dijk}}\ and\ \citenamefont
  {Nogami}(1999)}]{vandijk99}%
  \BibitemOpen
  \bibfield  {author} {\bibinfo {author} {\bibfnamefont {W.}~\bibnamefont {{van
  Dijk}}}\ and\ \bibinfo {author} {\bibfnamefont {Y.}~\bibnamefont {Nogami}},\
  }\href@noop {} {\bibfield  {journal} {\bibinfo  {journal} {Phys. Rev. Lett.}\
  }\textbf {\bibinfo {volume} {83}},\ \bibinfo {pages} {2867} (\bibinfo {year}
  {1999})}\BibitemShut {NoStop}%
\bibitem [{\citenamefont {{van Dijk}}\ and\ \citenamefont
  {Nogami}(2002)}]{vandijk02}%
  \BibitemOpen
  \bibfield  {author} {\bibinfo {author} {\bibfnamefont {W.}~\bibnamefont {{van
  Dijk}}}\ and\ \bibinfo {author} {\bibfnamefont {Y.}~\bibnamefont {Nogami}},\
  }\href@noop {} {\bibfield  {journal} {\bibinfo  {journal} {Phys. Rev. C}\
  }\textbf {\bibinfo {volume} {65}},\ \bibinfo {pages} {024608} (\bibinfo
  {year} {2002})}\BibitemShut {NoStop}%
\bibitem [{\citenamefont {Nogami}\ and\ \citenamefont {van
  Dijk}(2001)}]{nogami01}%
  \BibitemOpen
  \bibfield  {author} {\bibinfo {author} {\bibfnamefont {Y.}~\bibnamefont
  {Nogami}}\ and\ \bibinfo {author} {\bibfnamefont {W.}~\bibnamefont {van
  Dijk}},\ }\href@noop {} {\bibfield  {journal} {\bibinfo  {journal} {Few-Body
  Systems Supplement}\ }\textbf {\bibinfo {volume} {13}},\ \bibinfo {pages}
  {196} (\bibinfo {year} {2001})}\BibitemShut {NoStop}%
\bibitem [{\citenamefont {{van Dijk}}\ and\ \citenamefont
  {Nogami}(2003)}]{vandijk03a}%
  \BibitemOpen
  \bibfield  {author} {\bibinfo {author} {\bibfnamefont {W.}~\bibnamefont {{van
  Dijk}}}\ and\ \bibinfo {author} {\bibfnamefont {Y.}~\bibnamefont {Nogami}},\
  }\href@noop {} {\bibfield  {journal} {\bibinfo  {journal} {Few-Body Systems
  Suppl.}\ }\textbf {\bibinfo {volume} {14}},\ \bibinfo {pages} {229} (\bibinfo
  {year} {2003})}\BibitemShut {NoStop}%
\bibitem [{\citenamefont {Muller}(1999)}]{muller99a}%
  \BibitemOpen
  \bibfield  {author} {\bibinfo {author} {\bibfnamefont {H.~G.}\ \bibnamefont
  {Muller}},\ }\href@noop {} {\bibfield  {journal} {\bibinfo  {journal} {Phys.
  Rev. Lett.}\ }\textbf {\bibinfo {volume} {83}},\ \bibinfo {pages} {3158}
  (\bibinfo {year} {1999})}\BibitemShut {NoStop}%
\bibitem [{\citenamefont {Moyer}(2004)}]{moyer04}%
  \BibitemOpen
  \bibfield  {author} {\bibinfo {author} {\bibfnamefont {C.~A.}\ \bibnamefont
  {Moyer}},\ }\href@noop {} {\bibfield  {journal} {\bibinfo  {journal} {Am. J.
  Phys.}\ }\textbf {\bibinfo {volume} {72}},\ \bibinfo {pages} {351} (\bibinfo
  {year} {2004})}\BibitemShut {NoStop}%
\bibitem [{\citenamefont {Abramowitz}\ and\ \citenamefont
  {Stegun}(1965)}]{abramowitz65}%
  \BibitemOpen
  \bibfield  {author} {\bibinfo {author} {\bibfnamefont {M.}~\bibnamefont
  {Abramowitz}}\ and\ \bibinfo {author} {\bibfnamefont {I.~A.}\ \bibnamefont
  {Stegun}},\ }\href@noop {} {\emph {\bibinfo {title} {Handbook of Mathematical
  Functions}}}\ (\bibinfo  {publisher} {Dover Publications, Inc.},\ \bibinfo
  {address} {New York},\ \bibinfo {year} {1965})\BibitemShut {NoStop}%
\bibitem [{\citenamefont {Jeffreys}\ and\ \citenamefont
  {Jeffreys}(1946)}]{jeffreys46}%
  \BibitemOpen
  \bibfield  {author} {\bibinfo {author} {\bibfnamefont {H.}~\bibnamefont
  {Jeffreys}}\ and\ \bibinfo {author} {\bibfnamefont {B.~S.}\ \bibnamefont
  {Jeffreys}},\ }\href@noop {} {\emph {\bibinfo {title} {Methods of
  Mathematical Physics}}}\ (\bibinfo  {publisher} {Cambridge University
  Press},\ \bibinfo {address} {Cambridge},\ \bibinfo {year} {1946})\BibitemShut
  {NoStop}%
\bibitem [{\citenamefont {Puzynin}\ \emph {et~al.}(2000)\citenamefont
  {Puzynin}, \citenamefont {Selin},\ and\ \citenamefont
  {Vinitsky}}]{puzynin00}%
  \BibitemOpen
  \bibfield  {author} {\bibinfo {author} {\bibfnamefont {I.}~\bibnamefont
  {Puzynin}}, \bibinfo {author} {\bibfnamefont {A.}~\bibnamefont {Selin}}, \
  and\ \bibinfo {author} {\bibfnamefont {S.}~\bibnamefont {Vinitsky}},\
  }\href@noop {} {\bibfield  {journal} {\bibinfo  {journal} {Comp. Phys.
  Comm.}\ }\textbf {\bibinfo {volume} {126}},\ \bibinfo {pages} {158} (\bibinfo
  {year} {2000})}\BibitemShut {NoStop}%
\bibitem [{Note1()}]{Note1}%
  \BibitemOpen
  \bibinfo {note} {There is an error in Ref.~\cite {vandijk07}. The plus and
  minus signs in Eq.~(3.4) of that paper should be interchanged.}\BibitemShut
  {Stop}%
\bibitem [{\citenamefont {Shao}\ and\ \citenamefont {Wang}(2009)}]{shao09}%
  \BibitemOpen
  \bibfield  {author} {\bibinfo {author} {\bibfnamefont {H.}~\bibnamefont
  {Shao}}\ and\ \bibinfo {author} {\bibfnamefont {Z.}~\bibnamefont {Wang}},\
  }\href {\doibase 10.1103/PhysRevE.79.056705} {\bibfield  {journal} {\bibinfo
  {journal} {Phys. Rev. E}\ }\textbf {\bibinfo {volume} {79}},\ \bibinfo
  {pages} {056705} (\bibinfo {year} {2009})}\BibitemShut {NoStop}%
\bibitem [{\citenamefont {Peters}\ and\ \citenamefont
  {Maley}(1968)}]{peters68}%
  \BibitemOpen
  \bibfield  {author} {\bibinfo {author} {\bibfnamefont {G.~O.}\ \bibnamefont
  {Peters}}\ and\ \bibinfo {author} {\bibfnamefont {C.~E.}\ \bibnamefont
  {Maley}},\ }\href@noop {} {\bibfield  {journal} {\bibinfo  {journal} {Am.
  Math. Monthly}\ }\textbf {\bibinfo {volume} {75}},\ \bibinfo {pages} {741}
  (\bibinfo {year} {1968})}\BibitemShut {NoStop}%
\bibitem [{\citenamefont {St\"utzle}\ \emph {et~al.}(2005)\citenamefont
  {St\"utzle}, \citenamefont {G\"obel}, \citenamefont {H\"orner}, \citenamefont
  {Kierig}, \citenamefont {Mourachko}, \citenamefont {Oberthaler},
  \citenamefont {Efremov}, \citenamefont {Federov}, \citenamefont {Yakovlev},
  \citenamefont {{van Leeuwen}},\ and\ \citenamefont {Schleich}}]{stutzle05}%
  \BibitemOpen
  \bibfield  {author} {\bibinfo {author} {\bibfnamefont {R.}~\bibnamefont
  {St\"utzle}}, \bibinfo {author} {\bibfnamefont {M.~C.}\ \bibnamefont
  {G\"obel}}, \bibinfo {author} {\bibfnamefont {T.}~\bibnamefont {H\"orner}},
  \bibinfo {author} {\bibfnamefont {E.}~\bibnamefont {Kierig}}, \bibinfo
  {author} {\bibfnamefont {I.}~\bibnamefont {Mourachko}}, \bibinfo {author}
  {\bibfnamefont {M.~K.}\ \bibnamefont {Oberthaler}}, \bibinfo {author}
  {\bibfnamefont {M.~E.}\ \bibnamefont {Efremov}}, \bibinfo {author}
  {\bibfnamefont {M.~V.}\ \bibnamefont {Federov}}, \bibinfo {author}
  {\bibfnamefont {V.~P.}\ \bibnamefont {Yakovlev}}, \bibinfo {author}
  {\bibfnamefont {K.~A.~H.}\ \bibnamefont {{van Leeuwen}}}, \ and\ \bibinfo
  {author} {\bibfnamefont {W.~P.}\ \bibnamefont {Schleich}},\ }\href@noop {}
  {\bibfield  {journal} {\bibinfo  {journal} {Phys. Rev. Lett.}\ }\textbf
  {\bibinfo {volume} {95}},\ \bibinfo {pages} {110405} (\bibinfo {year}
  {2005})}\BibitemShut {NoStop}%
\bibitem [{\citenamefont {Berry}\ and\ \citenamefont {Balazs}(1979)}]{berry79}%
  \BibitemOpen
  \bibfield  {author} {\bibinfo {author} {\bibfnamefont {M.~V.}\ \bibnamefont
  {Berry}}\ and\ \bibinfo {author} {\bibfnamefont {N.~L.}\ \bibnamefont
  {Balazs}},\ }\href@noop {} {\bibfield  {journal} {\bibinfo  {journal} {Am. J.
  Phys.}\ }\textbf {\bibinfo {volume} {47}},\ \bibinfo {pages} {264} (\bibinfo
  {year} {1979})}\BibitemShut {NoStop}%
\bibitem [{\citenamefont {Besieris}\ \emph {et~al.}(1994)\citenamefont
  {Besieris}, \citenamefont {Shaarawi},\ and\ \citenamefont
  {Ziolkowski}}]{besieris94}%
  \BibitemOpen
  \bibfield  {author} {\bibinfo {author} {\bibfnamefont {I.~M.}\ \bibnamefont
  {Besieris}}, \bibinfo {author} {\bibfnamefont {A.~M.}\ \bibnamefont
  {Shaarawi}}, \ and\ \bibinfo {author} {\bibfnamefont {R.~W.}\ \bibnamefont
  {Ziolkowski}},\ }\href@noop {} {\bibfield  {journal} {\bibinfo  {journal}
  {Am. J. Phys.}\ }\textbf {\bibinfo {volume} {62}},\ \bibinfo {pages} {519}
  (\bibinfo {year} {1994})}\BibitemShut {NoStop}%
\bibitem [{\citenamefont {Durnin}\ and\ \citenamefont {{Miceli,
  Jr.}}(1987)}]{durnin87}%
  \BibitemOpen
  \bibfield  {author} {\bibinfo {author} {\bibfnamefont {J.}~\bibnamefont
  {Durnin}}\ and\ \bibinfo {author} {\bibfnamefont {J.~J.}\ \bibnamefont
  {{Miceli, Jr.}}},\ }\href@noop {} {\bibfield  {journal} {\bibinfo  {journal}
  {Phys. Rev. Lett.}\ }\textbf {\bibinfo {volume} {58}},\ \bibinfo {pages}
  {1499} (\bibinfo {year} {1987})}\BibitemShut {NoStop}%
\bibitem [{\citenamefont {Ryu}\ \emph {et~al.}(2014)\citenamefont {Ryu},
  \citenamefont {Henderson},\ and\ \citenamefont {Boshier}}]{ryu14}%
  \BibitemOpen
  \bibfield  {author} {\bibinfo {author} {\bibfnamefont {C.}~\bibnamefont
  {Ryu}}, \bibinfo {author} {\bibfnamefont {K.~C.}\ \bibnamefont {Henderson}}, \
  and\ \bibinfo {author} {\bibfnamefont {M.~G.}\ \bibnamefont {Boshier}},\
  }\href@noop {} {\bibfield  {journal} {\bibinfo  {journal} {New J. Phys.}\
  }\textbf {\bibinfo {volume} {16}},\ \bibinfo {pages} {013046} (\bibinfo
  {year} {2014})}\BibitemShut {NoStop}%
\bibitem [{\citenamefont {Goldberg}\ \emph {et~al.}(1967)\citenamefont
  {Goldberg}, \citenamefont {Schey},\ and\ \citenamefont
  {Swartz}}]{goldberg67}%
  \BibitemOpen
  \bibfield  {author} {\bibinfo {author} {\bibfnamefont {A.}~\bibnamefont
  {Goldberg}}, \bibinfo {author} {\bibfnamefont {H.~M.}\ \bibnamefont {Schey}},
  \ and\ \bibinfo {author} {\bibfnamefont {J.~L.}\ \bibnamefont {Swartz}},\
  }\href@noop {} {\bibfield  {journal} {\bibinfo  {journal} {Am. J. Phys.}\
  }\textbf {\bibinfo {volume} {35}},\ \bibinfo {pages} {177} (\bibinfo {year}
  {1967})}\BibitemShut {NoStop}%
\bibitem [{\citenamefont {Blanes}\ \emph {et~al.}(2009)\citenamefont {Blanes},
  \citenamefont {Casas}, \citenamefont {Oteo},\ and\ \citenamefont
  {Ros}}]{blanes09}%
  \BibitemOpen
  \bibfield  {author} {\bibinfo {author} {\bibfnamefont {S.}~\bibnamefont
  {Blanes}}, \bibinfo {author} {\bibfnamefont {F.}~\bibnamefont {Casas}},
  \bibinfo {author} {\bibfnamefont {J.}~\bibnamefont {Oteo}}, \ and\ \bibinfo
  {author} {\bibfnamefont {J.}~\bibnamefont {Ros}},\ }\href@noop {} {\bibfield
  {journal} {\bibinfo  {journal} {Phys. Rep.}\ }\textbf {\bibinfo {volume}
  {470}},\ \bibinfo {pages} {151} (\bibinfo {year} {2009})}\BibitemShut
  {NoStop}%
\bibitem [{\citenamefont {Kormann}\ \emph {et~al.}(2008)\citenamefont
  {Kormann}, \citenamefont {Holmgren},\ and\ \citenamefont
  {Karlsson}}]{kormann08}%
  \BibitemOpen
  \bibfield  {author} {\bibinfo {author} {\bibfnamefont {K.}~\bibnamefont
  {Kormann}}, \bibinfo {author} {\bibfnamefont {S.}~\bibnamefont {Holmgren}}, \
  and\ \bibinfo {author} {\bibfnamefont {H.~O.}\ \bibnamefont {Karlsson}},\
  }\href@noop {} {\bibfield  {journal} {\bibinfo  {journal} {J. Phys. Chem.}\
  }\textbf {\bibinfo {volume} {128}},\ \bibinfo {pages} {184101} (\bibinfo
  {year} {2008})}\BibitemShut {NoStop}%
\bibitem [{\citenamefont {Husimi}(1953)}]{husimi53}%
  \BibitemOpen
  \bibfield  {author} {\bibinfo {author} {\bibfnamefont {K.}~\bibnamefont
  {Husimi}},\ }\href@noop {} {\bibfield  {journal} {\bibinfo  {journal} {Prog.
  Theor. Phys.}\ }\textbf {\bibinfo {volume} {9}},\ \bibinfo {pages} {381}
  (\bibinfo {year} {1953})}\BibitemShut {NoStop}%
\bibitem [{\citenamefont {Moya-Cessa}\ and\ \citenamefont
  {Guasti}(2003)}]{moyacessa03}%
  \BibitemOpen
  \bibfield  {author} {\bibinfo {author} {\bibfnamefont {H.}~\bibnamefont
  {Moya-Cessa}}\ and\ \bibinfo {author} {\bibfnamefont {M.~F.}\ \bibnamefont
  {Guasti}},\ }\href@noop {} {\bibfield  {journal} {\bibinfo  {journal} {Phys.
  Lett. A}\ }\textbf {\bibinfo {volume} {311}},\ \bibinfo {pages} {1} (\bibinfo
  {year} {2003})}\BibitemShut {NoStop}%
\bibitem [{\citenamefont {Fityo}\ and\ \citenamefont
  {Tkachuk}(2005)}]{fityo05}%
  \BibitemOpen
  \bibfield  {author} {\bibinfo {author} {\bibfnamefont {F.}~\bibnamefont
  {Fityo}}\ and\ \bibinfo {author} {\bibfnamefont {V.}~\bibnamefont
  {Tkachuk}},\ }\href@noop {} {\bibfield  {journal} {\bibinfo  {journal}
  {Journal of Physical Studies}\ }\textbf {\bibinfo {volume} {9}},\ \bibinfo
  {pages} {299} (\bibinfo {year} {2005})}\BibitemShut {NoStop}%
\bibitem [{\citenamefont {Guedes}(2001)}]{guedes01}%
  \BibitemOpen
  \bibfield  {author} {\bibinfo {author} {\bibfnamefont {I.}~\bibnamefont
  {Guedes}},\ }\href@noop {} {\bibfield  {journal} {\bibinfo  {journal} {Phys.
  Rev. A}\ }\textbf {\bibinfo {volume} {63}},\ \bibinfo {pages} {034102}
  (\bibinfo {year} {2001})}\BibitemShut {NoStop}%
\bibitem [{\citenamefont {Houek}(2009)}]{houfek09}%
  \BibitemOpen
  \bibfield  {author} {\bibinfo {author} {\bibfnamefont {K.}~\bibnamefont
  {Houek}},\ }in\ \href@noop {} {\emph {\bibinfo {booktitle} {Numerical
  analysis and applied mathematics, International Conference 2009}}},\
  Vol.~\bibinfo {volume} {1},\ \bibinfo {editor} {edited by\ \bibinfo {editor}
  {\bibfnamefont {T.~E.}\ \bibnamefont {Simos}}, \bibinfo {editor}
  {\bibfnamefont {G.}~\bibnamefont {Psihoyios}}, \ and\ \bibinfo {editor}
  {\bibfnamefont {C.}~\bibnamefont {Tsitouras}}}\ (\bibinfo  {publisher}
  {American Institute of Physics},\ \bibinfo {year} {2009})\ pp.\ \bibinfo
  {pages} {293--296}\BibitemShut {NoStop}%
\bibitem [{Note2()}]{Note2}%
  \BibitemOpen
  \bibinfo {note} {Equation~(16) of Ref.~\cite {fityo05} has a typo: the 3
  should be replaced by 2.}\BibitemShut {Stop}%
\end{thebibliography}
%

\end{document}